\documentclass[journal]{IEEEtran}
\IEEEoverridecommandlockouts
\usepackage{cite}
\usepackage{amsmath,amssymb,amsfonts}
\usepackage{algorithm}
\usepackage{algorithmic}
\usepackage{graphicx}
\usepackage{textcomp}
\usepackage{xcolor}
\usepackage{bm}
\usepackage{caption}

\newtheorem{theorem}{Theorem}

\newtheorem{lemma}{Lemma}

\newtheorem{definition}{Definition}

\usepackage{comment}

\ifCLASSOPTIONcompsoc
  \usepackage[caption=false,font=normalsize,labelfont=sf,textfont=sf]{subfig}
\else
  \usepackage[caption=false,font=footnotesize]{subfig}
\fi

\makeatletter

\let\save@mathaccent\mathaccent
\newcommand*\if@single[3]{%
  \setbox0\hbox{${\mathaccent"0362{#1}}^H$}%
  \setbox2\hbox{${\mathaccent"0362{\kern0pt#1}}^H$}%
  \ifdim\ht0=\ht2 #3\else #2\fi
  }
\newcommand*\rel@kern[1]{\kern#1\dimexpr\macc@kerna}
\newcommand*\widebar[1]{\@ifnextchar^{{\wide@bar{#1}{0}}}{\wide@bar{#1}{1}}}
\newcommand*\wide@bar[2]{\if@single{#1}{\wide@bar@{#1}{#2}{1}}{\wide@bar@{#1}{#2}{2}}}
\newcommand*\wide@bar@[3]{%
  \begingroup
  \def\mathaccent##1##2{%
    \let\mathaccent\save@mathaccent
    \if#32 \let\macc@nucleus\first@char \fi
    \setbox\z@\hbox{$\macc@style{\macc@nucleus}_{}$}%
    \setbox\tw@\hbox{$\macc@style{\macc@nucleus}{}_{}$}%
    \dimen@\wd\tw@
    \advance\dimen@-\wd\z@
    \divide\dimen@ 3
    \@tempdima\wd\tw@
    \advance\@tempdima-\scriptspace
    \divide\@tempdima 10
    \advance\dimen@-\@tempdima
    \ifdim\dimen@>\z@ \dimen@0pt\fi
    \rel@kern{0.6}\kern-\dimen@
    \if#31
      \overline{\rel@kern{-0.6}\kern\dimen@\macc@nucleus\rel@kern{0.4}\kern\dimen@}%
      \advance\dimen@0.4\dimexpr\macc@kerna
      \let\final@kern#2%
      \ifdim\dimen@<\z@ \let\final@kern1\fi
      \if\final@kern1 \kern-\dimen@\fi
    \else
      \overline{\rel@kern{-0.6}\kern\dimen@#1}%
    \fi
  }%
  \macc@depth\@ne
  \let\math@bgroup\@empty \let\math@egroup\macc@set@skewchar
  \mathsurround\z@ \frozen@everymath{\mathgroup\macc@group\relax}%
  \macc@set@skewchar\relax
  \let\mathaccentV\macc@nested@a
  \if#31
    \macc@nested@a\relax111{#1}%
  \else
    \def\gobble@till@marker##1\endmarker{}%
    \futurelet\first@char\gobble@till@marker#1\endmarker
    \ifcat\noexpand\first@char A\else
      \def\first@char{}%
    \fi
    \macc@nested@a\relax111{\first@char}%
  \fi
  \endgroup
}
\makeatother

\begin{document}

%

\title{Peer Offloading in Mobile Edge Computing with Worst-Case Response Time Guarantees}
%
%
%

\author{Xingqiu~He,
        and~Sheng~Wang,~\IEEEmembership{Member,~IEEE}
\thanks{X. He and S. Wang are with the School of Communication and Information Engineering,
University of Electronic Science and Technology of China, Chengdu 610051, China (e-mail: hexqiu@gmail.com; wsh\_keylab@uestc.edu.cn).}
\thanks{Corresponding author: Sheng Wang.}
\thanks{This paper is accepted by IEEE Internet of Things Journal. Digital Object Identifier: 10.1109/JIOT.2020.3019492}
\thanks{Copyright (c) 2020 IEEE. Personal use of this material is permitted. However, permission to use this material for any other purposes must be obtained from the IEEE by sending a request to pubs-permissions@ieee.org.}}

\maketitle

\begin{abstract}
Mobile edge computing (MEC) is a new paradigm that provides cloud computing services at the edge of networks.
To achieve better performance with limited computing resources, peer offloading between cooperative edge servers (e.g. MEC-enabled base stations) has been proposed as 
an effective technique to handle bursty and spatially imbalanced arrival of computation tasks.
While various performance metrics of peer offloading policies have been considered in the literatures, 
the worst-case response time, a common Quality of Service(QoS) requirement in real-time applications, yet receives much less attention.
To fill the gap, we formulate the peer offloading problem based on a stochastic arrival model
and propose two online algorithms for cases with and without prior knowledge of task arrival rate. 
Our goal is to maximize the utility function of time-average throughput under constraints of energy consumption and worst-case response time.
Both theoretical analysis and numerical results show that our algorithms are able to produce close to optimal performance.
\end{abstract}


%
\IEEEpeerreviewmaketitle

\section{Introduction}
\label{section:introduction}
%
%
%
%
\IEEEPARstart{T}{he} ``pay-as-you-go'' cloud computing model has played a significant role in data storage and computation offloading in the past decade.
Recently, with the proliferation of smart devices and the development of Internet of Things, many new computationally intensive applications, such as smart cities and intelligent surveillance systems, have posed stringent quality of service (QoS) requirements that cloud computing is unable to meet.
To solve these problems and alleviate traffic congestions on transport networks, mobile edge computing (MEC) has emerged as a new paradigm to provide cloud computing services in close proximity to the end-users \cite{shi2016edge}.

Different from the traditional cloud computing framework where massive computing resources are placed on remote areas, MEC deploys computing servers throughout the network.
These servers are usually base stations (BSs), but can also be other dedicated devices with computing and storage resources.
Offloading computation tasks to nearby BSs rather than to the cloud substantially reduces end-to-end latency, thus improves the quality of experience (QoE) of end-users.
Extra tasks exceeding the computing capacity of local BSs are further offloaded to the cloud, forming a hierarchical offloading structure among end-users, BSs, and the cloud \cite{xu2018joint}.
Therefore, MEC is more like an extension rather than a substitute for cloud computing.
In addition to low-latency computing service, densely deployed BSs also provide other benefits like location awareness and mobility support.
Thus, MEC is considered as a promising approach to address the challenges posed by modern applications.

Although MEC is able to meet severe QoS requirements, one significant problem is that the available computing resources at the edge of the network are very limited compared to data centers in cloud computing.
Recently, peer offloading \cite{xiao2017qoe, chen2017computation, lyu2018distributed, li2019learning} has been proposed as an effective technique
to handle bursty and spatially imbalanced arrival of computation tasks.
By exploiting cooperation among BSs, peer offloading allows overloaded BSs to forward part of their workload to their neighbors,
thus improves the utilization of existing computing resources and user experience.

As far as we know, most existing researches \cite{xiao2017qoe, chen2017computation, lyu2018distributed} of peer offloading are based on the fluid-flow model.
They assume the workload of computation tasks is divisible and regard the task arrival process as a fluid-flow with a certain rate.
As a result, control algorithms based on the fluid-flow model only consider the expected arrival rate and ignore the variances of task arrivals.
If the actual arrival process is bursty, the amount of arrived tasks may be substantially larger than the average level in a short time interval.
In this case, the performance of these algorithms will degrade significantly and result in a large worst-case response time.
We present a simple example in Section \ref{section:known} to further illustrate our argument.
A similar discussion is also given in \cite{li2019learning} and they solve this problem by incorporating deadlines of tasks into decision-making.
However, the algorithm in \cite{li2019learning} is specially designed for computation-intensive tasks whose processing time ranges from minutes to hours or even days.
Moreover, although \cite{li2019learning} serves tasks with the best effort, they do not ensure all accepted tasks will be processed before their deadlines.
Therefore, there still lacks a peer offloading algorithm that is able to provide worst-case response time guarantees for
real-time applications who generally require the response time of tasks should be less than 100 milliseconds (ms) \cite{suzn2016delay, nunna2015enabling, gpp201922}.

In this paper, we formulate the peer offloading problem based on the stochastic arrival model.
Control decisions are made for individual tasks instead of abstracted task flows.
We deliver two efficient online algorithms that are able to yield close to optimal performance while providing worst-case response time bound.
The main contributions of our work are summarized as follows.

(1) We formalize the peer offloading problem in MEC networks based on the stochastic arrival model. The objective is to maximize the utility function of time-average throughput under a long-term energy consumption constraint and the worst-case response time requirement.
Our algorithms can be extended to include other time-average constraints easily.
To the best of our knowledge, we are the first that provides worst-case response time guarantees for real-time applications.

(2) We present a simple yet efficient algorithm when the expected arrival rate of computation tasks at each BS is known in advance. Theoretical analysis shows the algorithm is optimal both in system performance and response time.

(3) When the arrival rate is unknown, we develop an online algorithm that requires no prior information based on Lyapunov optimization. We show that the key subroutine of the algorithm is equivalent to the classical \emph{assignment problem}, and thus can be solved in $O(n^3)$ time \cite{papadimitriou1998combinatorial}. Theoretical analysis of the algorithm presents a $O(1/V)$-$O(V)$ tradeoff between system performance and worst-case response time bound, where $V$ is a tunable parameter.
We carry out extensive simulations with a real-world dataset to verify theoretical results and demonstrate that the proposed algorithm outperforms others under various settings.

The rest of this paper is organized as follows. In Section \ref{section:related}, we review related works in more detail. In Section \ref{section:system}, we present the system model and formalize the problem. In Section \ref{section:known}, we propose an optimal algorithm when the arrival rate of computation tasks is known. In Section \ref{section:unknown}, we develop an online algorithm based on Lyapunov optimization, and give related theoretical analysis. In Section \ref{section:practical}, several techniques are proposed to improve the practicality of our algorithms. In Section \ref{section:simulation}, numerical results are presented to demonstrate the performance of our algorithm. Section \ref{section:conclusion} concludes the paper and shows open problems for future work.

\section{Related Works}
\label{section:related}
The emerging MEC paradigm offers the possibility of supporting a large variety of new applications such as smart cities and intelligent surveillance systems \cite{li20185g, gpp201922}.
One of the main research points in MEC is the task offloading problem.
\cite{fan2018application, liu2017incentive, chen2018multi, chen2018socially, chen2018task} stand in the position of end-users to decide which task should be offloaded to nearby BSs in order to optimize objectives like latency\footnote{In the rest of this paper, we will use ``response time'' and ``latency'' interchangeably.} and energy consumption.
In contrast, we consider from the point of view of BSs and study how cooperative BSs can handle their tasks collaboratively to provide the best user experience.
Although collaborative computing is a common act in geographical load balancing originally proposed for data centers,
the main concern there is reducing operational cost with respect to spatial diversities of workload patterns \cite{liu2011greening} and electricity price differences across regions \cite{luo2015spatio}.
In contrast, we care about system performances like throughput and energy consumption in cooperative MEC.
Additionally, while the cooperative task offloading problem in MEC is online in nature, the problem considered in geographical load balancing is usually offline.
Therefore, techniques developed for geographical load balancing cannot be directly applied to MEC.

Recently, extensive researches have been conducted on the cooperation strategy between edge servers and incentive mechanism design \cite{tran2017collaborative, pang2017latency, sheng2015energy, cao2017joint, you2018energy, chen2017socially, lai2018optimal}.
The works closest to ours are those that design control algorithms for peer offloading \cite{xiao2017qoe, chen2017computation, lyu2018distributed, li2019learning}.
The work in \cite{xiao2017qoe} considers the users' QoE and the BSs' power efficiency in the MEC network.
They observe a fundamental tradeoff between these two metrics and develop a distributed optimization framework to achieve this tradeoff.
The authors in \cite{chen2017computation} present a framework for online computation peer offloading.
They theoretically characterize the optimal peer offloading strategy and show that the role of a computing server is determined by its pre-offloading marginal computation cost.
A distributed optimization for cost-effectiveness offloading decisions is considered in \cite{lyu2018distributed}.
All the three works aim to optimize the expected latency while the authors in \cite{li2019learning} discuss the necessity to consider the variability of response time.
To enhance satisfaction ratio, they incorporate deadlines of tasks into decision-making.
However, the algorithm in \cite{li2019learning} is specially designed for computation-intensive tasks whose processing time ranges from minutes to hours or even days.
In addition, they serve tasks in a best-effort way and do not offer any service level guarantees.
Although works in \cite{atapattu2020latency, lyu2018energy, deng2018workload, li2017efficient, zhu2018task} also adopt the stochastic arrival model and consider worst-case latency of computation tasks in MEC networks, they either investigate the user-to-BSs offloading problem, or only study control policies for a single BS.
Therefore, to the best of our knowledge, our research is the first work that presents peer offloading algorithms being able to provide worst-case service guarantees for real-time applications that generally require the response time be less than 100 ms \cite{suzn2016delay, nunna2015enabling, gpp201922}.

Lyapunov optimization is an online framework that solves time-average optimization problems.
The main advantage of this method is it can produce asymptotically optimal results without requiring prior knowledge of the system's random events.
Lyapunov optimization is extensively used to solve the offloading problem in MEC, including offloading between end-users \cite{qiao2019online, pu2016d2d, zhang2019near}, offloading from users to BSs \cite{mao2016dynamic, xu2018joint, jiang2015energy, huang2012dynamic, li2019lyapunov, liu2019dynamic}, and considering both situations simultaneously \cite{cui2019task, song2014energy}.
The peer offloading algorithms in \cite{chen2017computation, lyu2018distributed} are also based on Lyapunov optimization.
Most of these works either seek to reduce the task latency \cite{mao2016dynamic, xu2018joint, chen2017computation}, or aim for a minimized energy consumption \cite{cui2019task, qiao2019online, song2014energy, huang2012dynamic, liu2019dynamic, pu2016d2d}.
Due to the variety of practical problems, extra considerations are incorporated into the design of offloading algorithms, including energy harvesting \cite{qiao2019online, mao2016dynamic, zhang2019near}, service cache \cite{xu2018joint}, user mobility \cite{liu2019dynamic}, and cooperation incentives \cite{pu2016d2d, zhang2019near}.
However, all these works only concern time-average performance metrics.
To provide a worst-case latency guarantee, we have to significantly revise the traditional Lyapunov optimization method and formulate our problem based on a different arrival model.

\section{System Model}
\label{section:system}
\begin{figure}[!t]
    \centering
    \includegraphics[width=2.5in]{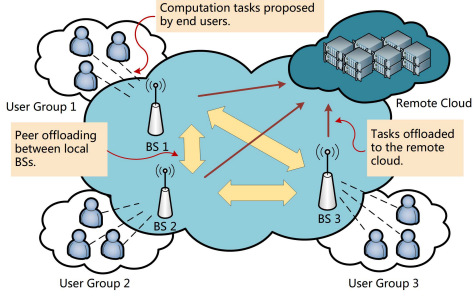}
    \caption{A simple MEC network with cooperative peer offloading.}
    \label{fig_fog1}
\end{figure}

We consider a local MEC network with $N$ BSs, which operates in slotted time $t$ as illustrated in Fig. \ref{fig_fog1}.
Similar to other peer offloading researches, we focus on the cooperation between BSs and does not consider the collaborative behaviors of end-users.
Our model is well suited for applications in which all tasks must be offloaded to edge servers,
either because end-users' devices lack adequate computation capabilities (e.g. sensors in IoT networks and cameras in intelligent surveillance systems),
or constrained by the energy budget (e.g. smartphones with a low battery).
In other applications, end-users may help each other to process computation tasks (e.g. offloading tasks from augmented reality glasses to a tablet or smartphone) and form a device-to-device (D2D) network.
In this case, one can integrate our work with scheduling policies in D2D networks \cite{kim2020joint} and let end-users choose offloading decisions with the minimal response time.
In our future work, we will also seek to extend our model to include user-side cooperation so that we can achieve the optimal performance with a unified scheduling control.

We assume that the total workload of all tasks arrived in every slot does not exceed the maximum workload that can be processed by
each BS in a single slot.
Since these tasks can be served within the same slot, we bind them as a whole and regard it as a single ``giant'' task.
We also assume temporarily that the workload of all ``giant'' tasks is equal.
In Section \ref{subsection:different_workload}, we show how to construct a general algorithm that is able to handle tasks of varying workload
from strategies designed in Section \ref{section:known} and \ref{section:unknown}.
In Section \ref{subsection:bounded_arrival}, we further demonstrate that our assumption should hold in most practical cases
and theoretically bound its impact if the assumption is not satisfied.

In each time slot, let $A_n(t)\in \{0,1\}$ be the random variable indicating whether there is
a task arrived at BS $n$.
Arrived tasks may be blocked if BSs are overloaded, and accepted tasks can be either processed locally or offloaded to nearby BSs.
We do not consider offloading tasks to the cloud because the round-trip time between the network edge and the remote cloud is
generally greater than the maximum latency allowed by real-time applications \cite{tomanek2016multidimensional}.
Besides, our work does not conflict with those that study the collaborative cloud and edge computing.
For example, one can integrate our algorithm with the framework proposed in \cite{ren2019collaborative} so that
(1) all latency-sensitive tasks are offloaded to BSs to get a delay guarantee and
(2) the rest tasks are allocated according to \cite{ren2019collaborative} to minimize the overall latency.
For the convenience of description, we temporarily assume all arrived tasks will be accepted and allow BSs to drop accepted tasks.
In Section \ref{subsection:block_decision}, we present a method that converts drop decisions to block decisions so that the refusal of service happens at the request stage and all accepted tasks are guaranteed to be served on time.

Let $\mu_n(t)$ and $D_n(t)$ be the amount of tasks processed and dropped by BS $n$ on time slot $t$.
$c_{nm}(t)$ is the number of tasks peer offloaded from BS $n$ to BS $m$.
Like $A_n(t)$, we require $\mu_n(t)$, $c_{mn}(t)$ and $D_n(t)$ are binary variables.
We use $\bm{Q}(t) = (Q_1(t),\dots,Q_N(t))$ to denote the number of tasks stored in the queues of BSs.
The update process of $Q_n(t)$ is
\begin{align}
	Q_n(t+1) =& \max [Q_n(t) - \mu_n(t) - \sum_{m\ne n}c_{nm}(t) - D_n(t) \notag \\
	&+ \sum_{m\ne n}c_{mn}(t-\delta_{mn}), 0] + A_n(t)
	\label{expr:Q_update_original}
\end{align}
where $\delta_{mn}$ is the one-way trip time from BS $m$ to BS $n$.
Thus $c_{mn}(t-\delta_{mn})$ is the number of peer offloaded tasks leaving BS $m$ on slot $t-\delta_{mn}$ and arriving at BS $n$ on slot $t$.

Our goal is to maximize the utility function of throughput with the constraint of time-average energy consumption and worst-case response time.
Standing in the position of BSs, the response time of a task in this paper refers to the time from the moment the task is received by BSs to the moment the computation result of the task is transmitted back to the user.
We omit the transmission time of the computation result as its size is usually very small.
Given the maximum latency $L^{max}$ allowed by users,
we want to solve the following stochastic optimization problem $P_o$ with the extra requirement that all non-dropped tasks
must be processed in $L^{max}$ time slots. The formulation of $P_o$ is
\begin{alignat}{2}
	\max\quad & \sum_{n} g_n(\widebar{y}_n) &{}& (P_o) \notag \\
	s.t.\quad & \widebar{Q}_n < \infty &\quad& \forall n\in\{1,\dots,N\} \notag  \\
	& \widebar{e}_n \leq E^{aver}_n &\quad& \forall n\in\{1,\dots,N\} \label{cons:po_power}
\end{alignat}
where 
\begin{gather}
    \widebar{y}_n \triangleq \lambda_n - \lim_{t\to\infty}\frac{1}{t}\sum^{t-1}_{\tau=0}\mathbb{E}[D_n(\tau)] \label{Expr:y-aver} \\
    \widebar{e}_n \triangleq \lim_{t\to\infty}\frac{1}{t}\sum^{t-1}_{\tau=0}\mathbb{E}[e_n(\tau)] \notag
\end{gather}
are the time-average expectation of throughput and energy consumption on BS $n$, respectively.
Here, $\lambda_n = \mathbb{E}[A_n(t)]$ is the expected task arrival rate of BS $n$ and $g_n$ is a concave function over $[0,1]$ that represents the utility of BS $n$.
Note that we have assumed a stationary $\lambda_n$ in order to simplify our statement,
but all algorithms and their performance analysis also hold when $\lambda_n$ is time-varying.
$E^{aver}_n$ is the upper bound of time-average energy consumption.
The energy consumption $e_n(t)$ depends on the computation activity $\mu_n(t)$.
Since $\mu_n(t)$ is binary, we use $e^1_n$ to denote the active energy consumption when $\mu_n(t)=1$ and
$e^0_n$ to denote the static energy consumption when $\mu_n(t)=0$.
Then we have $e_n(t) = e^1_n\mu_n(t) + e^0_n(1-\mu_n(t))$,
so the energy consumption constraint \eqref{cons:po_power} actually requires the time-average service level $\widebar{\mu}_n=\lim_{t\to\infty}{1}/{t}\sum^{t-1}_{\tau=0}\mathbb{E}[\mu_n(\tau)]$
satisfies $\widebar{\mu}_n \leq ({E^{aver}_n-e^0_n})/({e^1_n-e^0_n}$).

The difficulty of solving $P_o$ not only comes from the uncertainty of future task arrivals, but also from the coupling of decision variables
along the timeline. From \eqref{expr:Q_update_original} we can see that the state of $Q_n$ is dependent on the past peer offloading decisions $c_{mn}(t-\delta_{mn})$. 
To avoid this problem, we consider a relaxed problem $P_r$ where we set $\delta_{mn} = 0$ in $P_o$ for every $m,n \in \{1,2,\dots,N \}$.
Then, the update of $Q_n$ becomes
\begin{align}
	Q_n(t+1) =& \max [Q_n(t) - \mu_n(t) - \sum_{m\ne n}c_{nm}(t) - D_n(t) \notag \\
	&+ \sum_{m\ne n}c_{mn}(t), 0] + a_n(t)  \label{expr:Q_update_relaxed}.
\end{align}
The following theorem shows algorithms of $P_o$ can be constructed from algorithms of $P_r$.
\begin{theorem}
	If there is an algorithm $S^{*}_r$ for the relaxed problem $P_r$
	that achieves objective value $z^{*}_r$ with worst-case response time $T^{*}_r$, 
    then we can design an algorithm $S^{*}$ for the original problem $P_o$ that achieves $z^{*}_r$ with worst-case response time
	$T^{*}_r+2\delta^{max}$, where $\delta^{max} = \max_{mn}\delta_{mn}$.
    \label{theorem:1}
\end{theorem}
\begin{IEEEproof}
    To better describe the state change of $Q_n(t)$, we rewrite \eqref{expr:Q_update_relaxed} without the max operator
    \begin{align}
        Q_n(t+1) =& Q_n(t) - \widetilde{\mu}_n(t) - \sum_{m\ne n}\widetilde{c}_{nm}(t) - \widetilde{D}_n(t) \notag \\
        &+ \sum_{m\ne n}\widetilde{c}_{mn}(t) + a_n(t)  
        \label{expr:tilde_Q}
    \end{align}
    where $\widetilde{D}_n(t)$, $\widetilde{\mu}_n(t)$ and $\widetilde{c}_{mn}(t)$ are the actual number of tasks being dropped, being processed locally, and being peer offloaded, respectively.
    For example, if we have only one task in $Q_n(t)$ but $\mu_n(t)=1$ and $c_{nm}(t) = 1$ simultaneously.
    Since we cannot both offload and process this task, one of the above control decision must fail in execution.
    Thus, we have either $\widetilde{\mu}_n(t) = 0$ or $\widetilde{c}_{mn}(t) = 0$.
    One can prove that the time-average of control decisions and actual execution results are equal.
    The introduction of these notations are purely for the simplification of this proof.

    Since $\delta_{mn} = 0$ in $P_r$, the transmission of tasks is completed instantly. So there is no need to transmit tasks in advance and
    we can require that tasks are offloaded only when they will be served by other BSs in the next slot.
	Then, all tasks will be peer offloaded at most once.
	Let $(D^*_n(t),c^*_{mn}(t),\mu^*_n(t))$ and $(D^*_{r,n}(t),c^*_{r,mn}(t),\mu^*_{r,n}(t))$ be the decision variables of $S^*$ and $S^*_r$ respectively.
	For given $S^*_r$, let $D^*_n(t) = \widetilde{D}^*_{r,n}(t), c^*_{mn}(t) = \widetilde{c}^*_{r,mn}(t), \mu^*_n(t) = \widetilde{\mu}^*_{r,n}(t-\delta^{max})$.
    It is easy to check that $S^*$ is feasible for $P_o$. Next we focus on the performance of $S^*$.

	Since tasks can be peer offloaded at most once and the actual transmission time will not exceed $\delta^{max}$ slots, the task being served by BS $n$ on slot $t$ under $S^*_r$ is also available at BS $n$ on slot $t+\delta^{max}$ under $S^*$.
    Therefore, we have $\widetilde{\mu}_n^*(t) = \mu_n^*(t) = \widetilde{\mu}_{r,n}^*(t-\delta^{max})$. 
    This means tasks served on $t$ by $S_r^*$ will be served on $t+\delta^{max}$ by $S^*$. Thus the throughput, as well as the objective value, of $S^*$ is same to that of $S^*_r$. 
    Note that the computing result have to be transmitted back to the original BS, which cost no more than $\delta^{max}$ slots.
    Therefore, the worst-case response time of $S^{*}$ is $T^{*}_r+2\delta^{max}$.
\end{IEEEproof}

Theorem \ref{theorem:1} enables us to focus on algorithm design of $P_r$, which is a much easier problem because the update of $Q_n(t)$ no longer depends on
past decision variables.
In the next two sections, we design two online algorithms of $P_r$ for cases with and without prior information of task arrival rate.

\section{Algorithm Under Known Arrival Rate}
\label{section:known}
In this section, we assume the task arrival rate $\lambda_n$ is known.
We consider the following optimization problem $P_k$
\begin{alignat}{2}
    \max\quad & \sum_{n} g_n(\widehat{y}_n) &{}& (P_k) \label{problem:p-aux} \\
    s.t.\quad & 0\leq \widehat{\mu}_n \leq \frac{E^{aver}_n - e^0_n}{e^1_n - e^0_n} &\quad& \forall n\in \{1,\dots,N\} \label{P2-cons:power}\\
    & \widehat{y}_n \leq \lambda_n \label{P2-cons:arrival} &\quad& \forall n\in \{1,\dots,N\} \\
    & \sum_{n}\widehat{y}_n = \sum_{n}\widehat{\mu}_n &\quad& \forall n\in \{1,\dots,N\} \label{P2-cons:process}
\end{alignat}
where $\widehat{y}_n$ and $\widehat{\mu}_n$ are free variables in the set of real numbers.
Let $\widehat{\bm{y}}^* = (\widehat{y}^*_1, \dots, \widehat{y}^*_N),\widehat{\bm{\mu}}^* = (\widehat{\mu}^*_1, \dots, \widehat{\mu}^*_N)$ be the optimal solution of $P_k$ and $z^*$ be the corresponding optimal value.
The following theorem shows $z^*$ is an upper bound of system performance\footnote{The system performance here refers to the objective value of problem $P_o$.}.
\begin{theorem}
	No algorithm of $P_r$ can achieve an objective value greater than $z^*$.
\end{theorem}
\begin{IEEEproof}
	Suppose there is an algorithm $S'_r$ with objective value $z' > z^*$.
	Let $\widebar{y}_n'$ and $\widebar{\mu}_n'$ be the time-average throughput and service level of $S'_r$.
    The definition of of $\widebar{y}_n'$ \eqref{Expr:y-aver} implies $\widebar{y}_n'$ satisfies \eqref{P2-cons:arrival},
    and constraint \eqref{cons:po_power} implies $\widebar{\mu}_n'$ satisfies \eqref{P2-cons:power}.
    Summing \eqref{expr:tilde_Q} over $n$ results in
    \begin{equation}
        \sum_{n=1}^{N}Q_n(t) = \sum_{\tau=0}^{t-1}\sum_{n=1}^{N}A_n(\tau) - \sum_{\tau=0}^{t-1}\sum_{n=1}^{N}\widetilde{D}_n(\tau) - \sum_{\tau=0}^{t-1}\sum_{n=1}^{N}\widetilde{\mu}_n(\tau). 
        \label{eq:theorem2}
    \end{equation}
    Taking expectation, dividing by $t$, and letting $t\to \infty$.
    The left-hand side turns to $\lim_{t\to\infty}1/t \sum_{n=1}^{N}\mathbb{E}[Q_n(t)]$, which equals $0$ because $Q_n(t) < \infty$.
    Then \eqref{eq:theorem2} implies \eqref{P2-cons:process} by substituting \eqref{Expr:y-aver} into the right-hand side of \eqref{eq:theorem2}.
	Therefore, $\widebar{y}_n'$ and $\widebar{\mu}_n'$ are feasible variables of $P_k$ with objective value $z'$,
	contradicting the assumption that $z^*$ is the optimal value.
\end{IEEEproof}

From the above proof, we can see that
$\widehat{\bm{y}}^*$ and $\widehat{\bm{\mu}}^*$ are the time-average of optimal control decisions $y^*_n(t)$ and $\mu^*_n(t)$ of $P_r$.
Suppose the task arrival processes of different BSs are independent, we will show there is an algorithm that achieves
$z^*$ and serve all tasks within one slot.
The intuition behind the algorithm is illustrated by the following example.
Considering a 2 BSs MEC network with task arrival rate $(\lambda_1,\lambda_2) = (0.8,0.2)$ and energy consumption constraint that requires $(\widebar{\mu}_1,\widebar{\mu}_2) \leq (0.5,0.5)$.
Let $n_1, n_2$ denote the two BSs.
If we peer offload the task arrived at $n_1$ to $n_2$ with probability ${3}/{8}$, then the time-average number of tasks to be served by $n_1$ and $n_2$ are $\widebar{\mu}_1 = 0.8\times (1-{3}/{8}) = 0.5$ and
$\widebar{\mu}_2 = 0.2 + 0.8\times {3}/{8} = 0.5$, which satisfies the energy consumption constraint.
Note that such strategy is based on expected task arrival rate and is usually given by algorithms adopting the fluid-flow model.
We not show that although it achieves optimal throughput, the induced response time may be very large.
Assume on some slot $t$ we have $A_1(t) = 1$ and $A_2(t) = 1$, and we offload one task from $n_1$ to $n_2$.
Since the task arrival processes of different BSs are independent, such event happens with probability $0.2 \times 0.8\times {3}/{8} = 0.06$.
Because there are two tasks enter $Q_2(t)$ on slot $t$ and each BS can only process one task in every time slot,
one of the two tasks has to wait $1$ slot.
If in the next time slot, the same event happens again, then one of the four tasks has to wait $2$ slots.
Generally, for any finite integer $M$, there is a probability of at least $0.06^{M}$ that the response time of some tasks exceeds $M$ slots.

The problem of above strategy is that the control decisions only depend on the expected arrival rate and disregard the actual task arrival on each time slot.
As shown in the example, when the actual arrival differs from the expectation in a sequence of time slots, it inevitably induces a large response time.
In contrast, if we offload tasks of $n_1$ only when $A_2(t) = 0$,
then each BS is assigned at most one task on every slot and thus all newly arrived tasks can be served within one slot.
In our example, we first list the probabilities of all arrival events
\begin{align*}
    &p\left(A_1(t) = 0 \mbox{ and } A_2(t) = 0\right) = 0.2\times0.8 = 0.16 \\
    &p\left(A_1(t) = 1 \mbox{ and } A_2(t) = 0\right) = 0.8\times0.8 = 0.64 \\
    &p\left(A_1(t) = 0 \mbox{ and } A_2(t) = 1\right) = 0.2\times0.2 = 0.04 \\
    &p\left(A_1(t) = 1 \mbox{ and } A_2(t) = 1\right) = 0.8\times0.2 = 0.16
\end{align*}
Our strategy is offloading an arrived task from $n_1$ to $n_2$ with probability ${0.30}/{0.64}$ only when $A_1(t) = 1$ and $A_2(t) = 0$.
Then under all situations, there is at most one task enters the waiting queues of each BS so that all tasks can be served in the next slot.
The time-average service rate of $n_1$ is $\widebar{\mu}_1 = 0.64 \times \left( 1 - {0.30}/{0.64} \right) + 0.16 = 0.50$.
Similarly we can compute $\widebar{\mu}_2 = 0.50$.
So in this case both the throughput and the response time are optimal.
Now we extend this method to the general case.

Let $n_1,\dots,n_N$ denote the $N$ BSs.
Out goal is to compute how many tasks should be served by each BS given the actual arrival $\bm{A}(t) = (A_1(t), \dots, A_N(t))$.
We first decide how many tasks should be dropped so that the expected throughput equals $\widehat{\bm{y}}^*$.
In every slot $t$, observe $\bm{A}(t)$, then choose the value of $D_n(t)$ according to the following rule
\begin{equation}
    D_n(t) =
    \begin{cases}
        1& \text{with probability $1-{\widehat{y}^*_n}/{\lambda_n}$ when $A_n(t)=1$} \\
        0& \text{otherwise}
    \end{cases}
    \label{var:D}
\end{equation}
We use $\bm{A}^0(t) = (A^0_1(t), \dots, A^0_N(t))$ to denote the number of tasks accepted by local BSs, where $A^0_n(t) = A_n(t) - D_n(t)$.
It can be easily confirmed that for every $n$ and $t$, $A^0_n(t)$ is a $\{0,1\}$ random variable with expectation $\lambda_n - \lambda_n(1-{\widehat{y}^*_n}/{\lambda_n}) = \widehat{y}^*_n$.

Next, we develop a peer offloading strategy to let the time-average number of tasks processed by BSs equals $\widehat{\bm{\mu}}^*$.
The whole algorithm consists of $N$ steps.
In each step, we make offloading decisions based on the outcome of the previous step.
We use vector $\bm{A}^{i}(t)$ to denote both the output of $i$-th step and the input of $(n+1)$-th step.
The component $A^i_j(t)$ is the number of tasks assigned to BS $j$ by the end of step $i$.
The input of the first step is $\bm{A}^0(t)$.
Define operation $\pi_{ij}$ to swap the $i$-th and $j$-th component of any vector $\bm{A}$
\begin{align*}
    \pi_{ij}(A_1, \dots, A_i, &\dots, A_j, \dots, A_N) \\
    &= (A_1, \dots, A_j, \dots, A_i, \dots, A_N).
\end{align*}
For ease of statement, when the expectation of variables is invariant over time, their time index is omitted.
For example, we use $\mathbb{E}(A^{i-1}_i)$ instead of $\mathbb{E}(A^{i-1}_i(t))$.
Now we explain the $i$-th step of our algorithm in detail.
The overall procedure is summarized in Algorithm \ref{alg:known-arrival-rate}.

    (1) If $\mathbb{E}(A^{i-1}_i) = \widehat{\mu}^*_i$, let $\bm{A}^i(t) = \bm{A}^{i-1}(t)$ and skip to the next step.

    (2) Else, if $\mathbb{E}(A^{i-1}_i) < \widehat{\mu}^*_i$, it means the expected number of tasks assigned to $n_i$
    according to $\bm{A}^{i-1}$ is lower than $n_i$'s optimal time-average service rate, so
    we should assign more tasks to $n_i$ by offloading from other BSs.
    Find the smallest $m \in \{i, \dots, N\}$ such that
    \begin{equation}
        1 - (1-\mathbb{E}(A^{i-1}_i))\cdots(1-\mathbb{E}(A^{i-1}_m)) \geq \widehat{\mu}^*_i.
        \label{expr:choose_m}
    \end{equation}
    The left-hand side is the probability that there is at least one task arrived at $n_i,\dots,n_m$.
    Our strategy is offloading tasks arrived at the these BSs to $n_i$ so that the time-average number of tasks assigned to $n_i$ equals $\widehat{\mu}^*_i$.
    Specifically, in every time slot $t$, observe the value of $\bm{A}^{i-1}(t)$.
    If $A^{i-1}_i(t) = 1$, then no peer offloading is performed, and we have $\bm{A}^i(t) = \bm{A}^{i-1}(t)$.
    Else, find the smallest $p\in \{i+1,\dots,m\}$ such that $A^{i-1}_p(t) = 1$.
    If no such $p$ exists, let $\bm{A}^i(t) = \bm{A}^{i-1}(t)$.
    Otherwise, if $p<m$, then offload the task from $n_p$ to $n_i$.
    In this case, $\bm{A}^i(t) = \pi_{ip}(\bm{A}^{i-1}(t))$.
    If $p = m$, offload the task from $n_m$ to $n_i$ with probability
    \begin{equation}
    P_{m\to i} = \frac{\widehat{\mu}^*_i - [1-(1-\mathbb{E}(A^{i-1}_i))\cdots(1-\mathbb{E}(A^{i-1}_{m-1}))]}{(1-\mathbb{E}(A^{i-1}_i))\cdots(1-\mathbb{E}(A^{i-1}_{m-1}))\mathbb{E}(A^{i-1}_{m}))}.
        \label{expr:step1}
    \end{equation}
    Our choice of $m$ \eqref{expr:choose_m} guarantees that the value of \eqref{expr:step1} is non-negative.
    So when $p=m$ we have
    \begin{equation}
        \bm{A}^i(t) =
        \begin{cases}
            \pi_{im}(\bm{A}^{i-1}(t)) & \text{with probability $P_{m\to i}$} \\
            \bm{A}^{i-1}(t) & \text{with probability $1 - P_{m\to i}$}
        \end{cases}.
        \label{expr:choose_A_1}
    \end{equation}

    (3) Else, it must be $\mathbb{E}(A^{i-1}_i) > \widehat{\mu}^*_i$, we should offload tasks of $n_i$ to other BSs.
        Similarly, find the smallest $m \in \{i,\dots,N\}$ such that
    \begin{equation}
        \mathbb{E}(A^{i-1}_i)\mathbb{E}(A^{i-1}_{i+1})\dots\mathbb{E}(A^{i-1}_m) \leq \widehat{\mu}^*_i.
        \label{expr:choose_m_2}
    \end{equation}
    The left-hand side is the probability that there is a newly arrived task for all $n_i,\dots,n_m$.
    If $A^{i-1}_i(t)=1$ and $A^{i-1}_{i+1}(t)A^{i-1}_{i+2}(t)\cdots A^{i-1}_m(t)=0$, let $p$ be the least integer with $A^{i-1}_p(t) = 0$.
    Offload the task of $n_i$ to $n_p$ with probability
    \begin{equation*}
        P_{i\to p} = \frac{\mathbb{E}(A^{i-1}_i)-\widehat{\mu}^*_i}{\mathbb{E}(A^{i-1}_i)(1-\mathbb{E}(A^{i-1}_{i+1})\dots\mathbb{E}(A^{i-1}_m))}.
    \end{equation*}
    Likewise, this value must be non-negative.
    In this case
    \begin{equation}
        \bm{A}^i(t) =
        \begin{cases}
            \pi_{ip}(\bm{A}^{i-1}(t)) & \text{with probability $P_{i\to p}$} \\
            \bm{A}^{i-1}(t) & \text{with probability $1 - P_{i\to p}$}
        \end{cases}.
        \label{expr:choose_A_2}
    \end{equation}
    Otherwise, when $A^{i-1}_i(t)=0$ or $A^{i-1}_{i+1}(t)A^{i-1}_{i+2}(t)\cdots A^{i-1}_m(t)=1$, we have $\bm{A}^i(t) = \bm{A}^{i-1}(t)$.

\begin{algorithm}[t]
    \caption{Peer Offloading for Known Arrival Rate}
    \label{alg:known-arrival-rate}
    \begin{algorithmic}[1]
        \renewcommand{\algorithmicrequire}{\textbf{Input:}}
        \renewcommand{\algorithmicensure}{\textbf{Output:}}
        \REQUIRE Task arrival $\bm{A}(t)$, expected arrival rate $\bm{\lambda}$,
            optimal solution of problem \eqref{problem:p-aux} $(\widehat{\bm{y}}^*, \widehat{\bm{\mu}}^*)$
        \ENSURE  Offloading decision $\bm{A}^N(t)$
        \STATE Choose $\bm{D}(t)$ according to \eqref{var:D};
        \STATE $\bm{A}^0(t) \Leftarrow \bm{A}(t) - \bm{D}(t)$;
        \FOR{$i=1$ to $N$}
            \IF{$\mathbb{E}(A^{i-1}_{i}) = \widehat{\mu}^*_i$}
            \STATE $\bm{A}^i(t) \Leftarrow \bm{A}^{i-1}(t)$;
            \ELSIF{$\mathbb{E}(A^{i-1}_{i}) < \widehat{\mu}^*_i$}
                \STATE Find $m$ according to \eqref{expr:choose_m};
                \IF{$A^{i-1}_i(t) = 1$}
                    \STATE $\bm{A}^i(t) \Leftarrow \bm{A}^{i-1}(t)$;
                \ELSE
                    \STATE Find the smallest $p\in \{i+1,\dots,m\}$ such that $A^0_p(t) = 1$;
                    \IF{$p$ does not exist}
                        \STATE $\bm{A}^i(t) \Leftarrow \bm{A}^{i-1}(t)$;
                    \ELSIF{$p<m$}
                        \STATE $\bm{A}^i(t) \Leftarrow \pi_{ip}(\bm{A}^{i-1}(t))$;
                    \ELSE
                        \STATE Choose $\bm{A}^i(t)$ according to \eqref{expr:choose_A_1};
                    \ENDIF
                \ENDIF
            \ELSE
                \STATE Find $m$ according to \eqref{expr:choose_m_2};
                \IF{$A^{i-1}_i(t)=1$ AND $A^{i-1}_{i+1}(t)A^{i-1}_{i+2}(t)\cdots A^{i-1}_m(t)=0$}
                    \STATE Let $p\in \{i+1, \dots, m\}$ be the least integer with $A^{i-1}_p(t) = 0$;
                    \STATE Choose $\bm{A}^i(t)$ according to \eqref{expr:choose_A_2};
                \ELSE
                    \STATE $\bm{A}^i(t) \Leftarrow \bm{A}^{i-1}(t)$;
                \ENDIF
            \ENDIF
        \ENDFOR
        \RETURN $\bm{A}^N(t)$
    \end{algorithmic}
\end{algorithm}

Starting from the first step, one can verify that for each $i\in \{1,\dots,N\}$,
we have: (1) $A^i_j(t) \leq 1 \ \forall j\in \{1,\dots,N\}$; (2) $\sum_{j=1}^{N}A^i_j(t) = \sum_{j=1}^{N}A^{i-1}_j(t)$; (3) $\mathbb{E}(A^i_j(t)) = \widehat{\mu}^*_j \ \forall j\in \{1,\dots, i\}$.
Repeat the process $N$ times, it is guaranteed that the final output $\bm{A}^N(t)$ satisfies
\begin{gather}
    A^N_j(t) \leq 1  \quad  \forall j\in \{1,\dots,N\} \label{prop:AN_leqone} \\
    \sum_{j=1}^{N}A^N_j(t) = \sum_{j=1}^{N}A^0_j(t)  \label{prop:AN_equalsum} \\
    \mathbb{E}(A^N_j) = \widehat{\mu}^*_j \quad \forall j \in \{1,\dots,N\} \label{prop:AN_matchservice}
\end{gather}
Offload tasks so that the number of tasks assigned to each BS equals $\bm{A}^N(t)$ and let BSs serve the assigned tasks in the next slot.
The performance of the algorithm is analyzed as follows:
\begin{enumerate}
    \item Since we assign at most one task to each BS at every slot according to \eqref{prop:AN_leqone}, all non-dropped tasks will be served within one slot.
    \item Equation \eqref{prop:AN_matchservice} and constraint \eqref{P2-cons:power} guarantees the time-average energy consumption constraint is not violated, which means our algorithm is feasible.
    \item Since all non-dropped tasks are served by the $N$ BSs \eqref{prop:AN_equalsum},
        our choice of $\bm{D}(t)$ guarantees the throughput of all BSs equals $\widehat{\bm{y}}^*$, which produces optimal system performance $z^*$.
\end{enumerate}
Therefore, it can be concluded that our algorithm is optimal, both in system performance and response time.

It can be easily checked that the time complexity of Algorithm \ref{alg:known-arrival-rate} is $O(N^2)$.
One can also run the algorithm offline and store the output strategy for each possible arrival $\bm{A}(t)$.
This will consume $O(2^N)$ storage space in total.
After that, when the task arrival $\bm{A}(t)$ is observed, one can directly look up the corresponding offloading strategy without running the whole algorithm again.
The time complexity, in this case, is only $O(N)$.

\section{Algorithm Under Unknown Arrival Rate}
\label{section:unknown}
The optimality of the algorithm designed in the previous section largely depends on the prior knowledge of the arrival rate.
In this section, we will solve the problem without such prior knowledge based on a methodology of Lyapunov Optimization.
Different from traditional Lyapunov framework that only provides a time-average response time bound, we design a
virtual queue that enables us to bound the response time in the worst-case.
As stated in the proof of Theorem \ref{theorem:1}, we can assume that tasks are offloaded to other BSs only if they will be processed in the next slot
As a result, the decisions of peer offloading and task serving can be represented by a single variable.
Let $b^n_m(t)\in \{0,1\}$ be the number of tasks at BS $n$ that are offloaded to and served by BS $m$ on slot $t$.
Then $\eta_n(t) = \sum_{m\in N} b^n_m(t)$ is the number of tasks in $Q_n(t)$ being served on slot $t$.
Tasks offloaded to BS $m$ will be served immediately and will not enter $Q_m(t)$.
Now, the update of $Q(t)$ is 
\begin{equation*}
	Q_n(t+1) = \max[Q_n(t) - \eta_n(t) - D_n(t), 0] + A_n(t).
\end{equation*}
Considering the following constraints:
\begin{gather}
    0 \leq \eta_n(t) \leq 1 \label{cons:b_1} \\
    0 \leq \sum_{n\in N} b^n_m(t) \leq 1 \label{cons:b_2} \\
	0 \leq \eta_n(t) + D_n(t) \leq 1 \notag
\end{gather}
where all variables are binary.
The first two constraints require that, in every slot $t$, at most one task of $Q_n(t)$ can be served, and
each BS can serve at most one task.
The last constraint ensures that the number of tasks leaving $Q_n(t)$ is at most one, whether being served or being dropped.
We will see later that this constraint does not harm the optimal value and it is useful in transforming drop decisions into block decisions.

In the following subsections, we first transform our problem $P_r$ into an equivalent form.
Then we set a virtual queue to record the waiting time of the head-of-line task.
We define a drift function of queues and combine it with our objective function to form a drift-plus-penalty bound.
An algorithm is designed to minimize this bound.
Theoretical analysis shows that the algorithm presents a $O(1/V)$-$O(V)$ tradeoff between system performance and worst-case response time bound,
where $V$ is a tunable parameter.

\subsection{Problem Transformation}
Assume the right partial derivative of $g_n(y)$ over $[0,1]$ is bounded by a non-negative constant $\nu_n$.
Define the concave extension of $g_n(y)$ over $[-1,\infty)$ as
\begin{equation*}
	\widehat{g}_n(y) \triangleq g_n([y]^1_0) + \nu_n \min[y_n,0]
\end{equation*}
where $[y]^1_0 \triangleq \min [\max[y,0],1]$.
Clearly, $\widehat{g}_n(y)$ is non-decreasing, concave and $g_n(y) = \widehat{g}_n(y)$ when $\ 0\leq y \leq 1$.
We extend the objective function to allow variables of $g_n$ taking negative values.
This will be useful in bounding the response time.
For the sake of convenience, we also use $\widehat{g}(\bm{y})$ to denote $\sum_{n}\widehat{g}_n(y_n)$ in the following subsections.

With the extended objective function, we introduce a vector of auxiliary variables 
$\bm{\gamma}(t) = (\gamma_1(t),\dots,\gamma_N(t))$ to transform $P_r$ into the following problem $P_t$
\begin{alignat}{2}
	\max\quad & \widehat{g}(\widebar{\bm{\gamma}}) &{}& (P_t) \notag \\
	s.t.\quad & \widebar{y}_n \geq \widebar{\gamma}_n &\quad& \forall n\in \{1,\dots,N\} \label{cons:yr}  \\
    & -1 \leq \widebar{\gamma}_n \leq 1 &\quad& \forall n\in \{1,\dots,N\} \label{cons:gamma} \\
	& \widebar{e}_n \leq E^{aver}_n &\quad& \forall n\in \{1,\dots,N\} \label{cons:power-3}  \\
	& \widebar{Q}_n < \infty &\quad& \forall n\in \{1,\dots,N\} \notag
\end{alignat}

Note that one can always choose $\gamma_n(t) = y_n(t)$ to ensure \eqref{cons:yr} and \eqref{cons:gamma} are satisfied.
Since $\widehat{g}_n(y)$ is non-decreasing, the optimal solution of $\gamma_n(t)$ will make \eqref{cons:yr} holds with equality.
Recall that $\widehat{g}_n(y) = g_n(y)$ on $[0,1]$, $P_t$ and $P_r$ must have same optimal objective value.
Therefore, any algorithm solves $P_t$ also solves $P_r$.

To ensure constraint \eqref{cons:yr}, we introduce a virtual queue
\begin{equation}
	Z_n(t+1) = \max[Z_n(t) - \lambda_n + D_n(t) + \gamma_n(t), 0]
	\label{expr:queue_z}
\end{equation}
from which we have 
\begin{equation*}
	Z_n(t+1) \geq Z_n(t) - \lambda_n + D_n(t) + \gamma_n(t).
\end{equation*}
Summing over $\tau\in\{0,\dots,t-1\}$ and dividing by $t$ yields
\begin{equation*}
	\frac{Z_n(t) - Z_n(0)}{t} + \frac{1}{t}\sum_{\tau =0}^{t-1}(\lambda_n - D_n(\tau)) \geq \frac{1}{t}\sum_{\tau=0}^{t-1}\gamma_n(\tau).
\end{equation*}
Take expectations of both sides and substituting $Z_n(0) = 0$, we have
\begin{equation}
    \frac{\mathbb{E}[Z_n(t)]}{t} + \widebar{y}_n(t) \geq \widebar{\gamma}_n(t).
	\label{expr:z_virtual_queue}
\end{equation}
It is apparent that when the virtual queue is stabilized, which means ${\mathbb{E}[Z_n(t)]}/{t} \to 0$ as $t\to \infty$, 
then the constraint \eqref{cons:yr} is satisfied.
Similarly, we introduce another virtual queue for constraint \eqref{cons:power-3}
\begin{equation*}
	W_n(t+1) = \max[W_n(t) - E^{aver}_n + e_n(t), 0].
\end{equation*}

It should be noted that the implementation of the virtual queue $Z_n(t)$ requires the knowledge of task arrival rate $\lambda_n$, which contradicts
the assumption that $\lambda_n$ is unknown.
Our plan is to temporarily assume $\lambda_n$ is known and develop an algorithm with performance analysis.
Later, in Section \ref{subsection:back}, we will replace $\lambda_n$ with past observation of task arrival $A_n(t)$, and show 
that the performance analysis still holds with slight modification.

\subsection{Waiting Time Virtual Queue}
In order to bound the maximum response time, we follow the technique used in \cite{neely2013delay}
and design a virtual queue $H_n(t)$ to record the waiting time of the head-of-line task in $Q_n(t)$.
Since all tasks will be in the head-of-line position before they are processed,
we bound the waiting time of all tasks in $Q_n(t)$ if we bound the length of $H_n(t)$, and thus we also bound the response time.
Set $H_n(t) = 0$ when $Q_n(t)$ is empty.
Define $\alpha_n(t)$ as an indicator variable that is $1$ if $Q_n(t) > 0$, and $0$ if the queue is empty.
Let $\beta_n(t) = 1 - \alpha_n(t)$.
The update rule of $H_n(t)$ is
\begin{align*}
    H_n(t\!+\!1) \!= &\alpha_n(t)\max[H_n(t) \!+\! 1 \!-\! (\eta_n(t)\!+\!D_n(t))T_n(t), 0] \notag \\
    &+ \beta_n(t)A_n(t)
\end{align*}
where $T_n(t)$ represents the inter-arrival time between the head-of-line task and the subsequent task.
The value of $T_n(t)$ is unknown if the subsequent task has not arrived yet.
Because arrivals are Bernoulli, if $H_n(t) > 0$, then $T_n(t)$ is a geometric random variable with success probability $\lambda_n$. 
If $H_n(t) = 0$, then we define $T_n(t) = 0$.

Without loss of generality, assume that $\lambda_n > 0$ for all BS $n\in \{ 1,2,\dots,N \}$.
Define $\bm{\Theta}(t) \triangleq [\bm{Z}(t);\bm{W}(t);\bm{H}(t)]$ and the following Lyapunov function
\begin{equation*}
	L(\bm{\Theta}(t)) \triangleq \frac{1}{2}\sum_{n=1}^{N}Z_n(t)^2 + \frac{1}{2}\sum_{n=1}^{N}W_n(t)^2 + \frac{1}{2}\sum_{n=1}^{N}\lambda_nH_n(t)^2.
	\label{expr:lya_func}
\end{equation*}
We now apply the Lyapunov optimization to develop an algorithm with bounded response time.

\subsection{Drift-Plus-Penalty Bound}
Define the one-step conditional Lyapunov drift
\begin{equation}
    \Delta(\bm{\Theta}(t)) \triangleq \mathbb{E}[L(\bm{\Theta}(t+1)) - L(\bm{\Theta}(t)) | \bm{\Theta}(t)].
    \label{expr:drift}
\end{equation}
Intuitively, $\Delta(\bm{\Theta}(t))$ describes the change of length of queues.
Recall that our goal is to maximize the objective function while bounding the length of queues.
Therefore, we can put $\Delta(\bm{\Theta}(t))$ and the objective function together and 
try to minimize them on every time slot.
Specifically, we form the following ``drift-plus-penalty'' term with parameter $V$ that decides the performance-latency tradeoff
\begin{equation}
    \Delta(\bm{\Theta}(t)) - V\mathbb{E}[\widehat{g}(\bm{\gamma}(t)) | \bm{\Theta}(t)].
	\label{expr:obj_queue}
\end{equation}

Before deducing the bound of \eqref{expr:obj_queue}, we introduce an independence property that will be useful in the proof.
\begin{definition}
	An algorithm has the independence property if for any slot $t$, every BS $n$ with 
	$H_n(t) > 0$ has a value of $T_n(t)$ that is independent of $\bm{\Theta}(t)$, $\eta_n(t)$, and $D_n(t)$.
\end{definition}

Since the arrivals are independent over queues and i.i.d. over slots, all algorithms that make decisions up to time $t$ independent of 
$T_n(t)$ have the independence property.
\begin{lemma}
	On every slot $t$, for any value of $\bm{\Theta}(t)$, and under any control policy that satisfies the independence property, we have
	\begin{align}
        \Delta(\bm{\Theta}(t)&) - V\mathbb{E}[\widehat{g}(\bm{\gamma}(t)) | \bm{\Theta}(t)] \leq B - V\mathbb{E}[\widehat{g}(\bm{\gamma}(t)) | \bm{\Theta}(t)] \notag \\
        & - \sum_{n}W_n(t)\mathbb{E}[E^{aver}_n-e_n(t) | \bm{\Theta}(t) ] \notag \\
        & - \sum_{n}Z_n(t)\mathbb{E}[\lambda_n-D_n(t)-\gamma_n(t) | \bm{\Theta}(t) ] \notag \\
        & - \sum_{n}H_n(t)\mathbb{E}[\eta_n(t)+D_n(t)-\lambda_n | \bm{\Theta}(t) ]   \label{lemma1-eq1}
	\end{align}
	where $B$ is a constant defined in the proof.
	\label{lemma:drift_bound_1}
\end{lemma}
\begin{IEEEproof}
The proof is given in Appendix \ref{appendix:1}.
\end{IEEEproof}

In the next subsection, we design an algorithm to minimize the right-hand side of \eqref{lemma1-eq1}.

\subsection{Algorithm Design} \label{section:algorithm}
Leaving out all constant terms in the right-hand side of \eqref{lemma1-eq1},
then minimizing the bound equals to maximizing the following expression
\begin{align*}
    &V\mathbb{E}[\widehat{g}(\bm{\gamma}(t)) | \bm{\Theta}(t)] - \sum_{n}Z_n(t)\mathbb{E}[D_n(t)+\gamma_n(t) | \bm{\Theta}(t) ] \\
    \!-\! &\sum_{n}W_n(t)\mathbb{E}[e_n(t) | \bm{\Theta}(t) ] \!+\! \sum_{n}H_n(t)\mathbb{E}[\eta_n(t)\!+\!D_n(t) | \bm{\Theta}(t) ] .
\end{align*}
The $\gamma_n(t)$ terms are separated from other decision variables, so we can optimize them separately by maximizing 
$V\widehat{g}(\bm{\gamma}) - \sum_{n}Z_n(t)\gamma_n(t)$, based on the observed $\bm{\Theta}(t)$ and subject to the constraint
$-1 \leq \gamma_n(t) \leq 1$.
After this, we are left with
\begin{equation}
	\sum_{n}H_n(t)\eta_n(t) + \sum_{n}(H_n(t)-Z_n(t))D_n(t) - \sum_{n}W_n(t)e_n(t).
	\label{expr:secondControlTerm}
\end{equation}
Since both $e_n(t)$ and $\eta_n(t)$ are related to $b^n_m(t)$ and we have the constraint
$D_n(t)+\eta_n(t) \leq 1$, all variables are correlated.
Clearly, in order to maximize \eqref{expr:secondControlTerm}, $D_n(t)$ should take value $1$ when $\eta_n(t) = 0$ and $H_n(t) \geq Z_n(t)$ and $Q_n(t) > 0$.
Otherwise $D_n(t) = 0$.
Define
\begin{equation}
	m_n(t) = 
	\begin{cases}
		1,& \text{if $Q_n(t)>0$ and $H_n(t)\geq Z_n(t)$} \\
		0,& \text{otherwise}
	\end{cases}.
	\label{expr:def_m}
\end{equation}
Then, we have $D_n(t) = m_n(t)(1-\eta_n(t))$.
Substitute into \eqref{expr:secondControlTerm} and leave out the constant term $\sum_{n}H_n(t)m_n(t)$ yields
\begin{equation}
	\sum_{n}\eta_n(t)[H_n(t)-m_n(t)(H_n(t)-Z_n(t))] - \sum_{n}W_n(t)e_n(t). \label{expr:remain_bound}
\end{equation}
From \eqref{expr:def_m}, we have
\begin{equation*}
	H_n(t)-m_n(t)(H_n(t)-Z_n(t)) = \min[H_n(t), Z_n(t)].
\end{equation*}
This is because if $m_n(t) = 0$, then $H_n(t)\leq Z_n(t)$, and so $H_n(t) = \min[H_n(t),Z_n(t)]$.
If $m_n(t) = 1$, then $H_n(t) \geq Z_n(t)$, and $Z_n(t) = \min[H_n(t), Z_n(t)]$.
Substituting into \eqref{expr:remain_bound} results in
\begin{equation*}
    \sum_{n}\min[H_n(t),Z_n(t)]\eta_n(t) - \sum_{n}W_n(t)e_n(t).
\end{equation*}
Replacing $\eta_n(t)$ and $e_n(t)$ with $\sum_{m}b^n_m(t)$ and $(e^1_n - e^0_n)\sum_{m}b^m_n(t)+e^0_n$,
leaving out the constant term, and exchanging the summing order, we finally have
\begin{equation}
	\max \sum_{n}\sum_{m}b^n_m(t)\left( \min[H_n(t),Z_n(t)] - W_m(t)(e^1_m-e^0_m) \right)
	\label{prob:schedule}
\end{equation}
subject to \eqref{cons:b_1} and \eqref{cons:b_2}.
It can be easily proved that this problem is equivalent to the well-known assignment problem by setting all negative coefficients
of $b^n_m$ to $0$ and forcing the left-hand of inequality constraints equals $1$.
We omit the detailed proof due to space limitation.
Therefore, our problem has the same time complexity with the assignment problem and thus can be solved in $O(n^3)$ time \cite{papadimitriou1998combinatorial}.
Putting everything together, our algorithm is summarized as follows.
On every slot $t$, observe $\bm{Z}(t)$, $\bm{W}(t)$, and $\bm{H}(t)$, then
\begin{enumerate}
	\item Choose $\bm{\gamma}(t) = (\gamma_1(t), \dots,\gamma_N(t))$ as the solution to the following problem:
\begin{align*}
	\max:& \quad V\widehat{g}(\bm{\gamma}(t)) - \sum_{n}Z_n(t)\gamma_n(t) \\
	s.t.:& -1 \leq \gamma_n(t) \leq 1 \quad \forall n\in \{1,\dots,N\}.
\end{align*}
Since our utility function is separable, this problem can be decomposed into $N$ single-variable problems, which 
have a closed-form solution when the concave function $\widehat{g}_n(\gamma_n)$ has a derivative.

\item Observe $\bm{Z}(t)$, $\bm{W}(t)$, $\bm{H}(t)$ and choose $b^n_m(t)$ to solve the optimization problem
\eqref{prob:schedule}.

\item For any BS $n$ with $\eta_n(t) = 0$, drop the head-of-line task if the queue is not
empty and $H_n(t) \geq Z_n(t)$.

\item Update all queues with variable values decided in the previous stages.
\end{enumerate}

\subsection{Performance Analysis} \label{subsection:performance}
From the description of our algorithm, we can see that the decisions made up to slot $t$ are independent of the value of $T_n(t)$, which indicates our algorithm possesses
the independence property.
Define $H^{max}_{n}$ for each BS $n$
\begin{equation*}
	H^{max}_{n} \triangleq \lceil V\nu_n \rceil + 2.
\end{equation*}
Let $H^{max}_g = \max_n[H^{max}_n]$.
The following theorem states that our algorithm presents a $O(1/V)$-$O(V)$ tradeoff between system performance and worst-case response time bound.
\begin{theorem}
	Suppose all queues are initially empty. The gap between the achieved system performance of our algorithm
	and the optimal value is bounded by $B/V$
	\begin{equation}
		\liminf_{t\to \infty} g(\widebar{\bm{y}}(t)) \geq g^* - B/V
		\label{expr:utility_bound}
	\end{equation}
	where $g^*$ is the optimal value of $P_r$.
	Meanwhile, we have a bound for all queues
	\begin{gather*}
		Q_n(t) \leq H_n(t) \leq H^{max}_n \\
		Z_n(t) \leq H^{max}_n \\
		W_n(t) \leq \lceil H^{max}_g/(e^1_n-e^0_n) \rceil + e^1_n - E^{aver}_n.
	\end{gather*}
    Since $H_n(t)$ records the waiting time of tasks in BS $n$, we also bound the worst-case response time.
	\label{theorem:performance}
\end{theorem}

Different from the common Lyapunov optimization framework which only provides a time-average response time bound, our algorithm considers the worst case.
By recalling Theorem \ref{theorem:1}, to ensure the worst-case response time is less than or equal to the given bound $L^{max}$, we only need
to choose a proper timescale for each slot $t$ and set the value of $V$ so that $\lceil V\max_n\{\nu_n\} \rceil + 2 \leq L^{max} - 2\delta^{max}$.
Based on our experience, the recommended timescale of each slot is around $1/50$ of the worst-case latency required by applications, but can also be flexibly adjusted to suit practical situations.
Some applications may have extra requirements, such as their tasks must be served by BSs that cached related databases/libraries.
These requirements can be transformed into constraints of decision variables.
This may increase the complexity of each step, but the algorithm structure and corresponding performance analysis are not influenced.
The proof of Theorem \ref{theorem:performance} is given in Appendix \ref{appendix:2}.

\subsection{Back to Unknown Arrival Rate}
\label{subsection:back}
Recall in \eqref{expr:queue_z}, the update of $Z_n(t)$ requires the value of $\lambda_n$.
Now we will fix this problem by replacing $\lambda_n$ with the past observation of the task arrival.
Particularly, we will use the following update rule of $Z_n(t)$
\begin{equation*}
	Z_n(t+1) = \max[Z_n(t) - A_n(t-W) + D_n(t) + \gamma_n(t), 0]
\end{equation*}
where the constant $W$ is equal to $H^{max}_{g}$. 
We can follow the similar way as in the previous subsection to prove that Theorem \ref{theorem:performance} still holds with 
$B$ replaced by a new term $B'+4W$, where $B'$ is a constant derived similarly as $B$.
The choice of $W$ guarantees $A_n(t-W)$ is independent with the current system state $\bm{\Theta}(t)$, which will be useful 
in bounding the quadratic term in drift-plus-penalty.
The detailed proof is omitted due to space limitations.

\section{More Practical Algorithms}
\label{section:practical}
In previous sections, we have assumed: (1) all arrived tasks are accepted and we allow BSs to drop accepted tasks;
(2) the workload of different tasks is the same.
In this section, we present methods to revise our algorithms so that they can better fit in real-world situations
where the above assumptions generally do not hold.

\subsection{Early Refuse}
\label{subsection:block_decision}
In common practice, tasks are not expected to be dropped once they are accepted.
The refusal of service usually should happen at an early stage where users propose new requests to nearby BSs.
These requests may be blocked if BSs are overloaded.
In this subsection, we show how to transform drop decisions $D_n(t)$ into block decisions.
Note that in Algorithm \ref{alg:known-arrival-rate}, the value of $D_n(t)$ is decided on the same slot when tasks arrive,
so the drop decisions there can be directly regarded as block decisions.
Hence, we only need to consider the algorithm designed in Section \ref{section:unknown}.

According to the 3-rd step of the algorithm in Section \ref{section:unknown}, the head-of-line task of BS $n$ is dropped only if $\eta_n(t)=0$.
By recalling that $\eta_i(t) \leq 1 \ \forall i \in \{ 1,\dots, N\}$, we know that $\sum_i \eta_n(t) \leq N-1$,
so there is at least one BS, denoted as BS $m$, which does not process any tasks on slot $t$.
Our technique is to let BS $m$ serve the head-of-line task of BS $n$, and on the same time block the next task arrived at BS $n$.
If we denote the head-of-line task as $\tau^n_1$ and the next arrived task as $\tau^n_2$.
Suppose according to the original algorithm, $\tau^n_1$ is dropped on slot $t_1$, and $\tau^n_2$ is served by BS $m'$ on slot $t_2$.
Then what we did can be understood as swapping $\tau^n_1$ and $\tau^n_2$ and shifting the process of $\tau^n_2$ from $t_2$ to $t_1$.
By exchanging the service order of these two tasks, the refusal of service is brought forward to an early stage.
If the task $\tau^n_2$ is dropped by the original algorithm, we only need to block yet another arrived task.

The only problem is that if $m\ne m'$, then we have
assigned an extra task to BS $m$, which may violate its energy consumption constraint.
To solve this, we only need to compensate BS $m$ by letting BS $m'$ help process a task originally assigned to BS $m$.
It is apparent that the response time, system throughput, and energy consumption are not changed after applying our technique
so the performance analysis in previous sections still holds, but now we have guaranteed that all accepted tasks will be served by local BSs.

\subsection{Tasks with Different Workload}
\label{subsection:different_workload}
The algorithms proposed in previous sections are specially designed for cases where computation tasks have an equal workload.
Such an assumption rarely holds in practice as tasks offloaded from users usually belong to different applications.
To improve the practicality of our algorithms, we
partition the range of workload into $K$ intervals $[l_k, u_k] \ \forall k\in \{1,\dots,K\}$, and
classify computation tasks into $K$ classes according to the interval their workload lies in.
Then we can construct $K$ instances of our algorithms to handle tasks in different classes.
The range of different intervals is usually equal, but can also be set to distinct values so that
the number of tasks contained in each class is approximately the same.
By increasing the number of classes, we can narrow down each interval and be more close to the equal workload assumption.
However, this is usually unnecessary in practice.
In fact, although we made the equal workload assumption when designing our algorithm, it is mainly for the tractability of theoretical analysis.
In realistic situations, our algorithm functions well even if this assumption is not satisfied.
Thus, there is no need to set $K$ to a large value.
In our experiments, the improvement is very limited when $K$ is greater than $5$.

For an arbitrary BS $n$, one can imagine instances of algorithms as several virtual machines running on top of it.
In every slot, each instance makes its own control decisions.
If it decides to serve a task (either from its own backlog or offloaded from other BSs), it sends the task to the substrate BS for processing.
The only resource of BS $n$ shared by the $K$ instances of algorithms
is CPU cycles constrained by the maximum time-average energy consumption $E^{aver}_n$.
Consequently, we have to divide the time-average process capacity of BS $n$ into $K$ parts.
This is done by computing an energy consumption constraint $E^{aver}_{n,k}$ for each instance $k$,
where $E^{aver}_{n,k}$ satisfies $\sum_k E^{aver}_{n,k} \leq E^{aver}_n$.
The value of each $E^{aver}_{n,k}$ depends on our goal.
For example, if we seek to maintain fairness,
we can re-assign $E^{aver}_{n,k}$ every a few slots proportional to the time-average workload of different classes computed from the history of task arrivals.
Our allocation is guaranteed to converge to the optimal value if the arrival process is ergodic.

The above assignment constrains the time-average number of tasks sent by upper instances.
It ensures these tasks can be processed by substrate BSs without violating BSs' energy constraints.
However, this assurance only holds in the long run.
In a specific time slot, the computing capacity may be insufficient if multiple instances decide to process a task simultaneously.
For example, assume the CPU frequency of BS $n$ is $f_n$, and the duration of each slot is $T$,
then $f_nT$ is the maximum number of CPU cycles that can be served by BS $n$ in one slot.
If $\sum_k u_k \leq f_nT$, then the process requests of all instances can be realized
even if they decide to serve a task at BS $n$ simultaneously.
On the other hand, if $\sum_k u_k > f_nT$, then the process of some tasks may have to be delayed due to the insufficiency of CPU power.
In this case, there is a possibility that the response time of these delayed tasks exceeds the bound derived in Section \ref{section:unknown}.
However, we can alleviate this problem by
using a smaller $V$ so that there is a gap between the derived bound $H^{max}_n$ and $L^{max}$ and leave more time to process
delayed tasks.



\subsection{Impact of the Maximum Arrival Assumption}
\label{subsection:bounded_arrival}
In Section \ref{section:system}, we assumed the arrived workload in each slot should not exceed the maximum workload that can be served
by BSs in a single slot.
In this subsection, we will first show that this assumption should hold in most practical cases.
After that, we derive a theorem to bound its impact to our algorithm if this assumption is not satisfied.

As shown in \cite{ko2018wireless}, in a widely used model, every BS is associated with a group of $N_u$ users, and each of them generates
a task request with probability $p_{u}$ in each time slot.
As a result, the number of requests received by BSs, denoted by $n_r$, follows a binomial distribution with expectation $N_up_u$.
If the timescale of each slot is short, $p_u$ is usually quite small so the probability of receiving $n_r$ requests decrease exponentially as $n_r$ becomes larger.
Therefore, in practical situations, our assumption will hold with a large probability if there is a gap between the time-average arrived workload
and the BS's peak process capacity.
For example, suppose the workload of each task is $w$ and let $w^{max}$ be the maximum workload that can be served by each BSs in a single slot.
If $N_u=100$, $p_u=0.1$, $w^{max}=15w$, then our assumption holds in more than $96\%$ cases.

Now we will demonstrate that our algorithm is barely affected even if this assumption is not satisfied.
We begin by defining two kinds of time regions.
Considering an arbitrary BS $n$.
An \emph{edge region} is a time interval such that $H_n(t) \geq H^{max}_n$ for all slot $t$ in that interval.
The definition of a \emph{violation region} is similar except that we require $H_n(t) \geq H^{max}_n + T$
for some constant $T$ in each slot of the region.
For arbitrary slot $t$, let $p_e$ and $p_v$ be the probability that $t$ belongs to an edge region and a violation region respectively.
If $p_{v|e}$ is the conditional probability that a time slot in the edge region also belongs to a violation region, then we have
$p_v = p_e p_{v|e}$ and the following theorem.
\begin{theorem}
    For all $T\geq3$, we have $p_{v|e} \leq M(r)/r^T$ where $r$ is a tunable parameter and $M(r)$ is a constant associated with $r$.
\end{theorem}
\begin{IEEEproof}
    The proof is given in Appendix \ref{appendix:3}.
\end{IEEEproof}
The above theorem demonstrates that $p_{v|e}$, as well as $p_v$, decreases exponentially as $T$ grows larger,
let alone to mention that $p_e$ itself is very small in practice.
Consequently, even if our assumption does not hold, we can still ensure tasks will be served in time by leaving a margin between
$H^{max}_n$ and the maximum latency allowed by users.
For example, if the worst-case response time bound required by users is 50 time slots, then we can choose
$V$ so that $H^{max}_n = 45$ or $40$, leaving a $5$ to $10$ slots gap.
We are guaranteed that the users' requirement will be satisfied with an extremely large probability.

We also want to mention that the data used in our numerical experiments do not meet the maximum arrival assumption,
but our algorithm still behaves as expected and outperforms other algorithms under various settings.
Therefore, both the theoretical analysis and experimental results validate the practicality of our algorithm.

\section{Simulation}
\label{section:simulation}

\begin{figure*}[!t]
\begin{minipage}[t]{0.49\linewidth}
\centering
\subfloat[Response Time]{\includegraphics[width=1.7in]{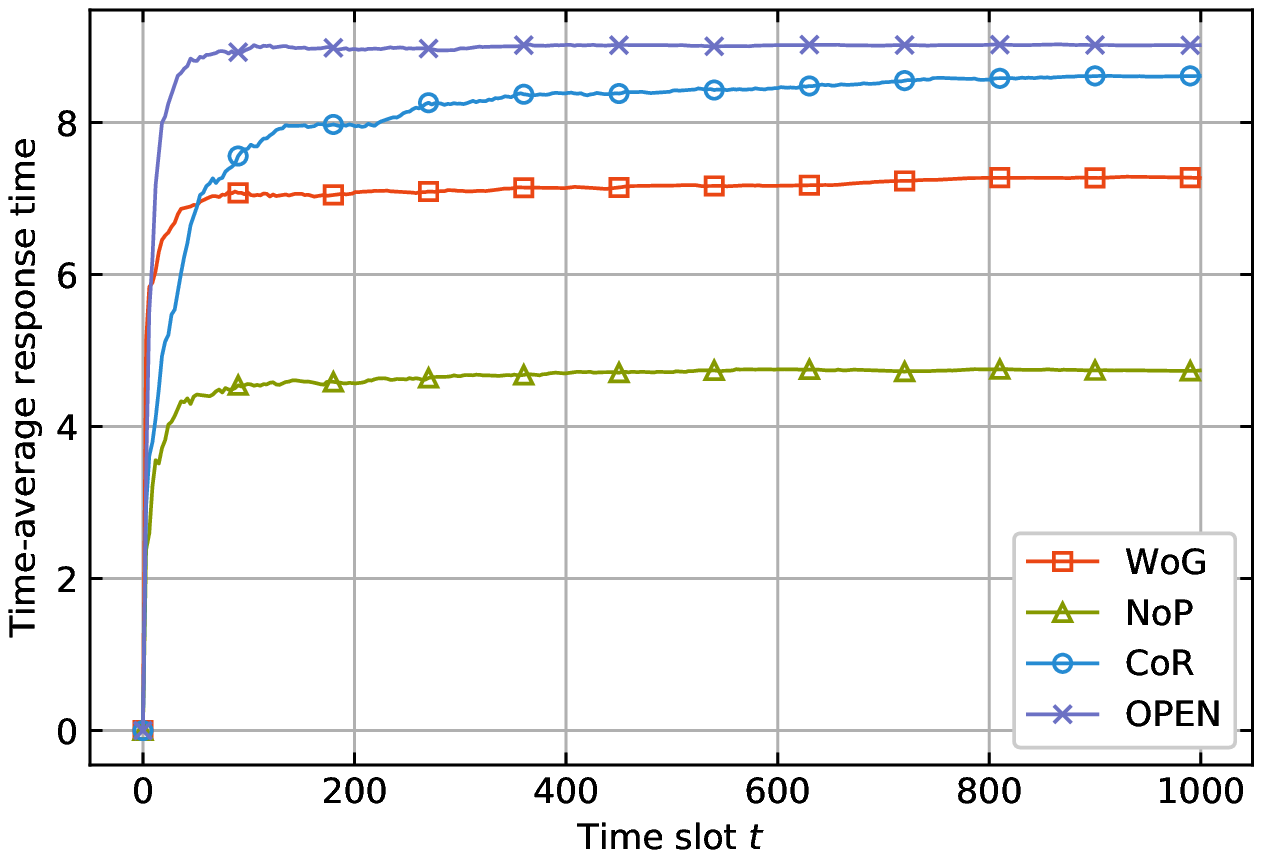} \label{light_load_delay}}
\hfil
\subfloat[System Utility]{\includegraphics[width=1.7in]{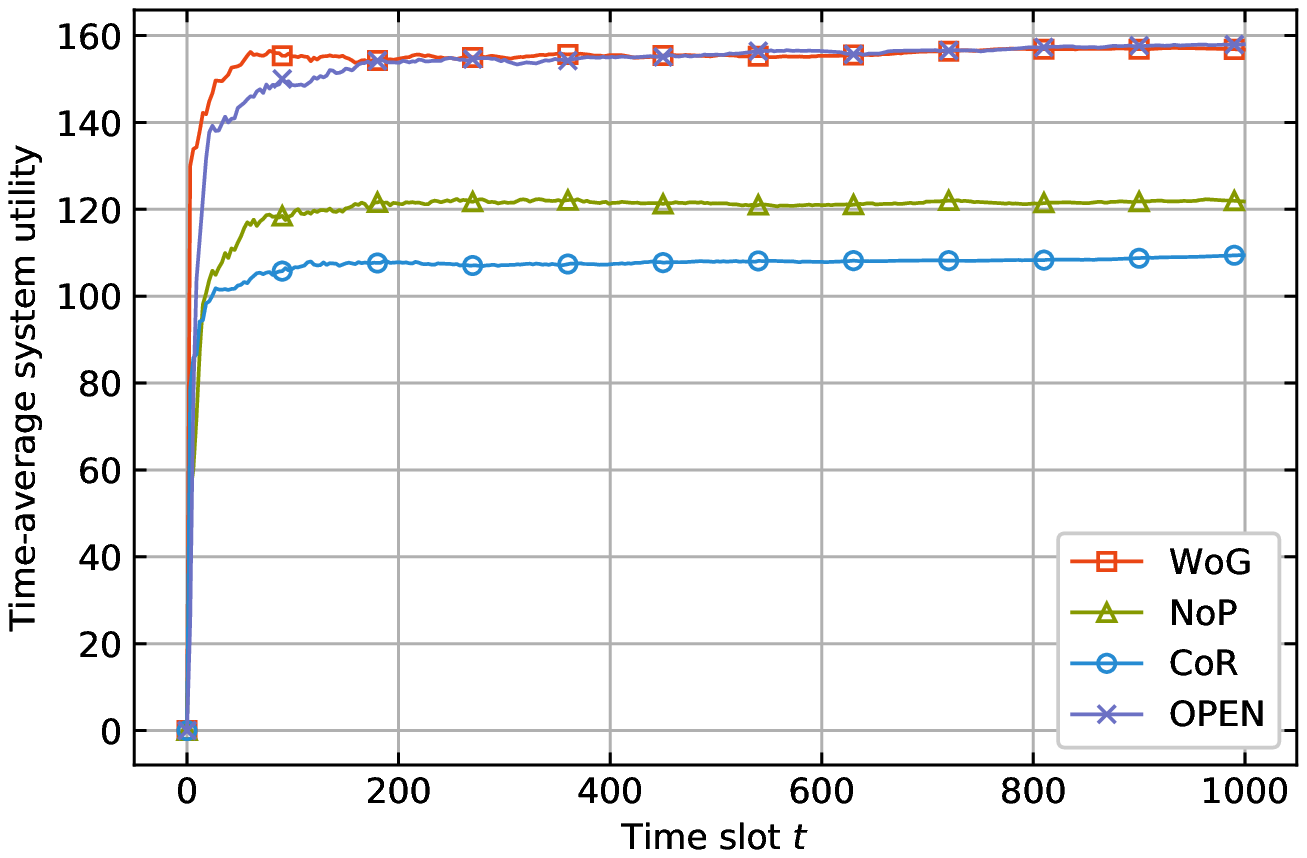} \label{light_load_utility}}
\caption{Time-average performance.}
\label{light_load}
\end{minipage}
\begin{minipage}[t]{0.49\linewidth}
\centering
\subfloat[Response Time]{\includegraphics[width=1.7in]{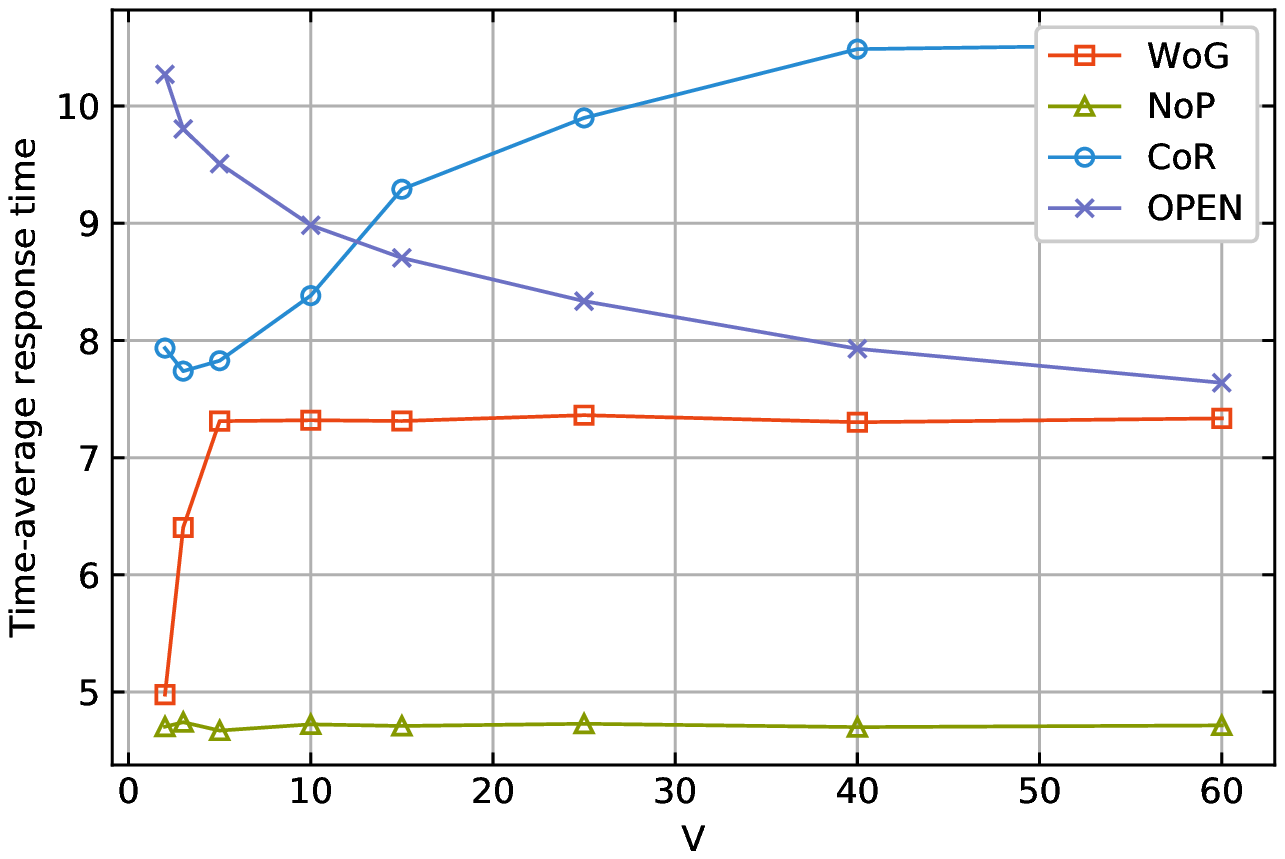} \label{V_delay}}
\hfil
\subfloat[System Utility]{\includegraphics[width=1.7in]{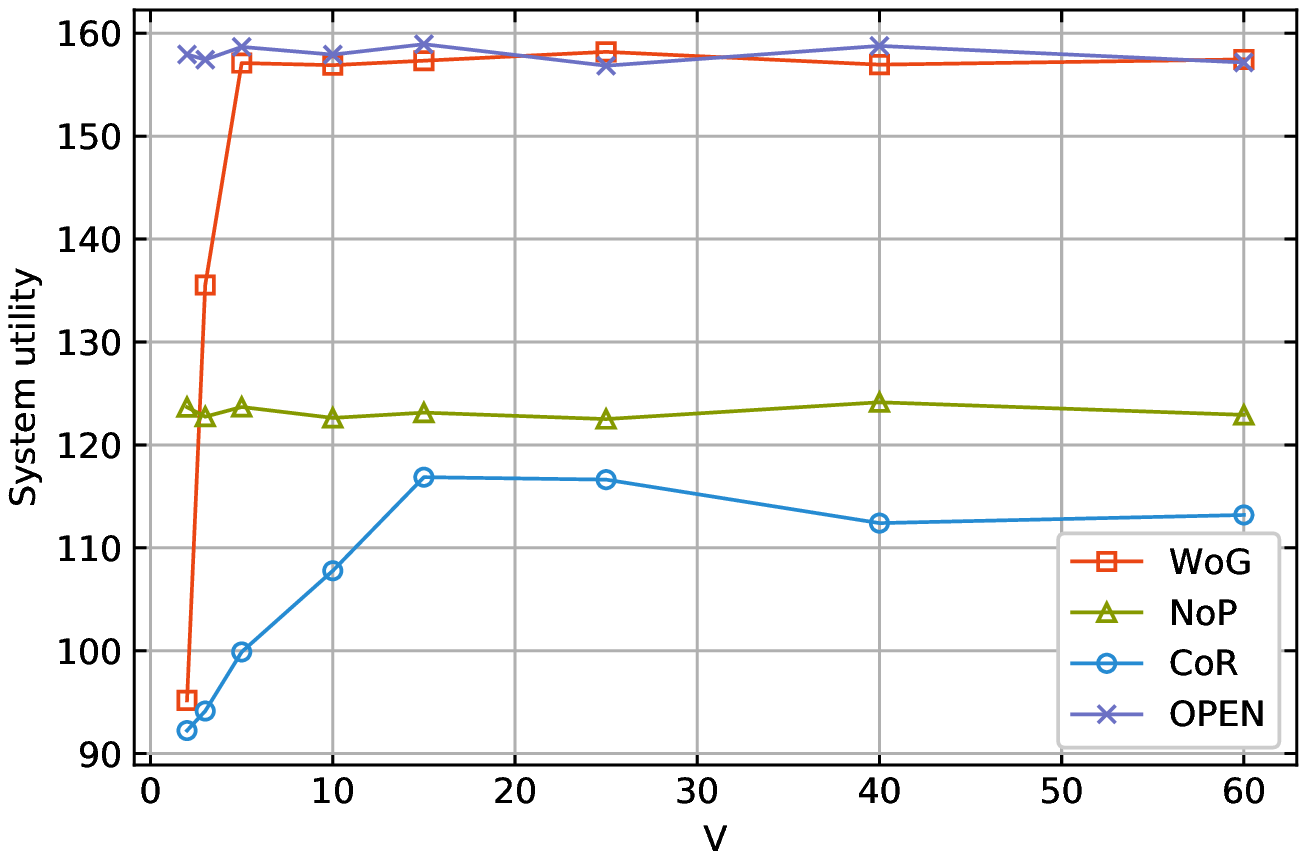} \label{V_utility}}
\caption{Impact of parameter V.}
\label{diff_V}
\end{minipage}
\end{figure*}

\begin{figure*}[!t]
\centering
\subfloat[Block Rate]{\includegraphics[width=1.7in]{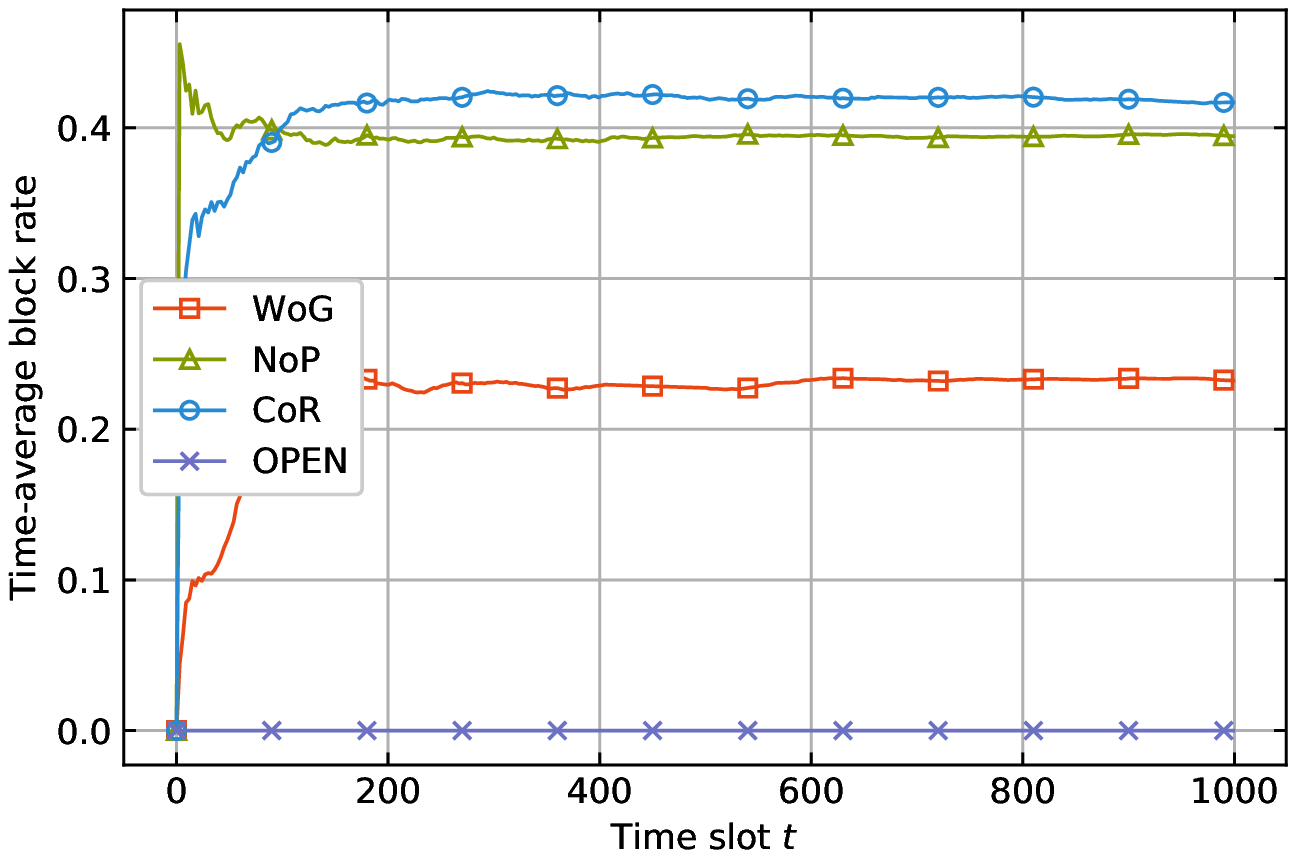} \label{heavy_load_block_rate}}
\hfil
\subfloat[Satisfaction Ratio]{\includegraphics[width=1.7in]{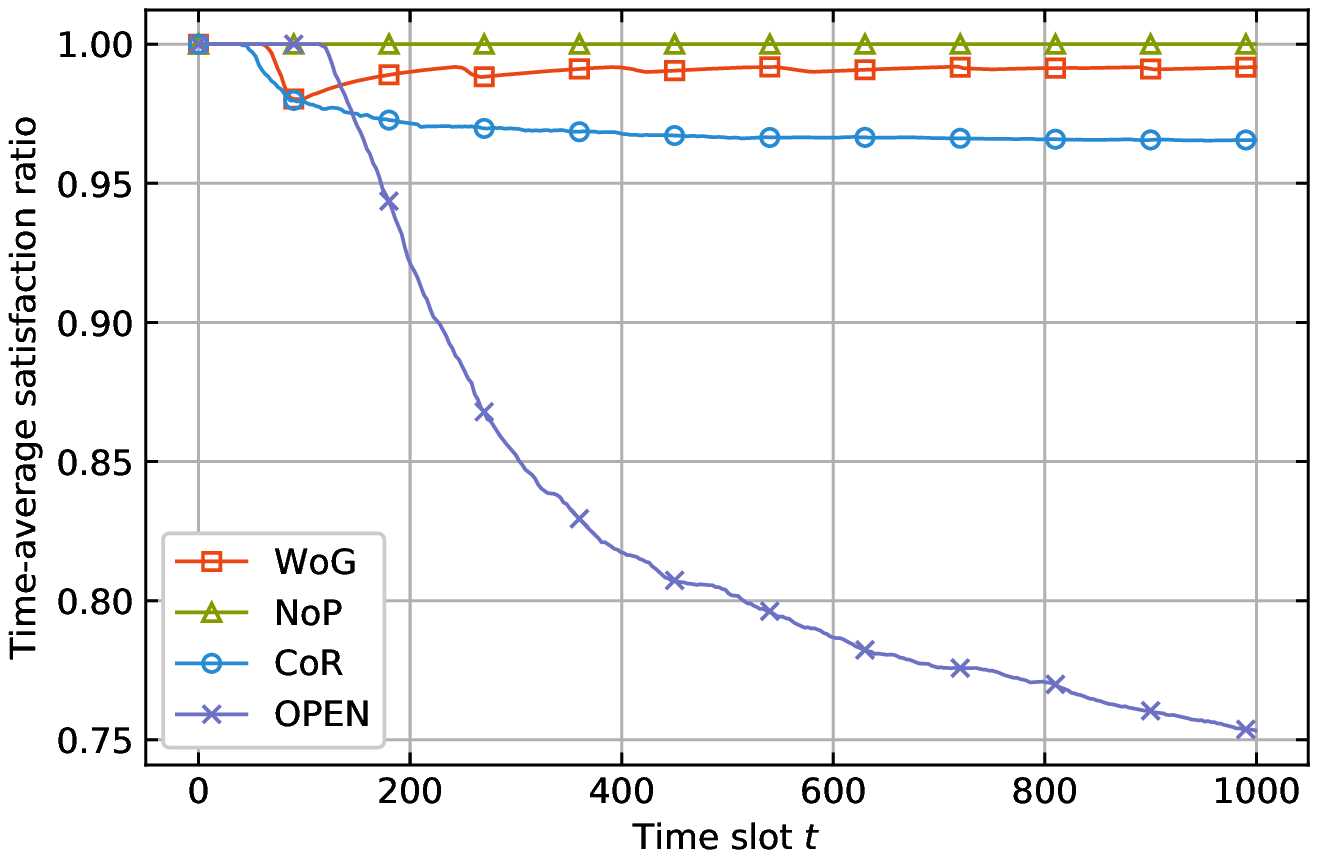} \label{heavy_load_satis_ratio}}
\hfil
\subfloat[Response Time]{\includegraphics[width=1.7in]{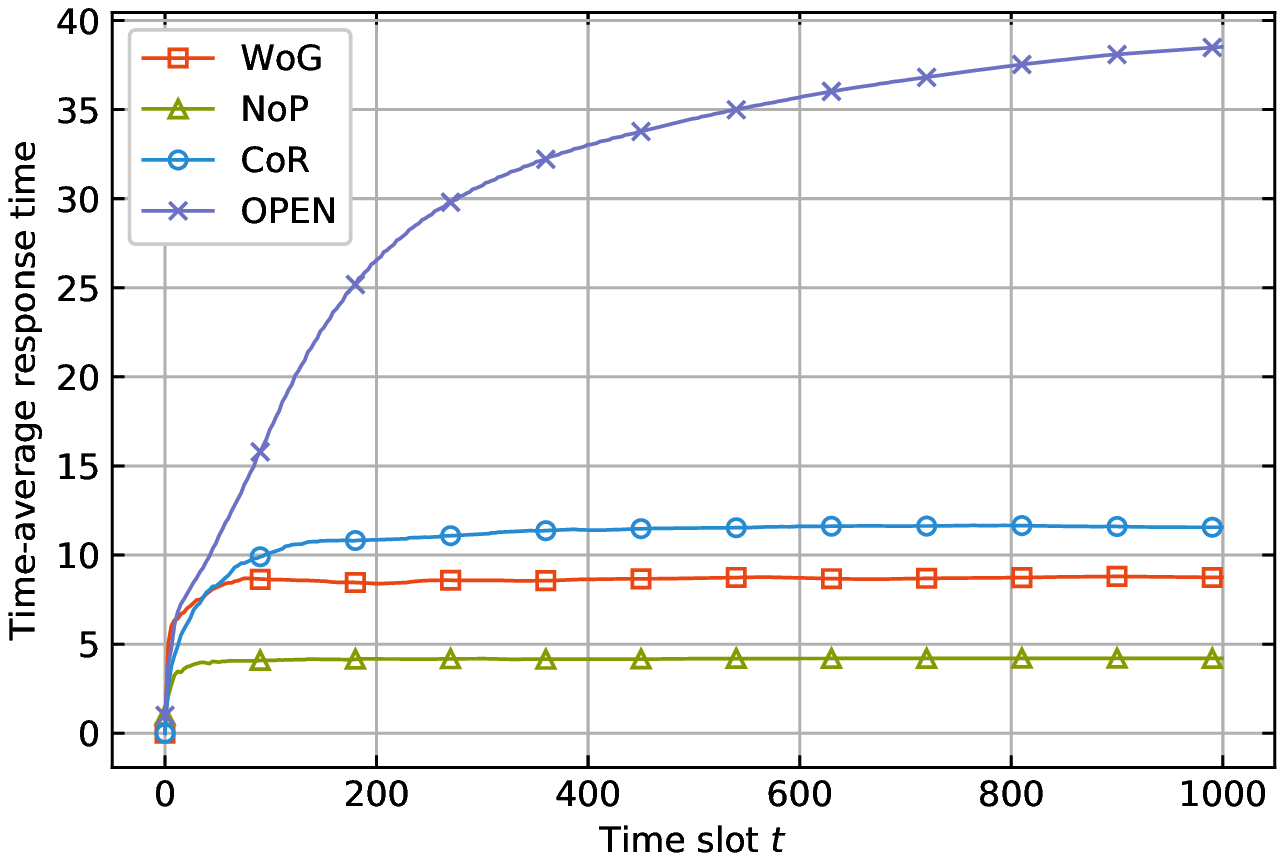} \label{heavy_load_delay}}
\hfil
\subfloat[System Utility]{\includegraphics[width=1.7in]{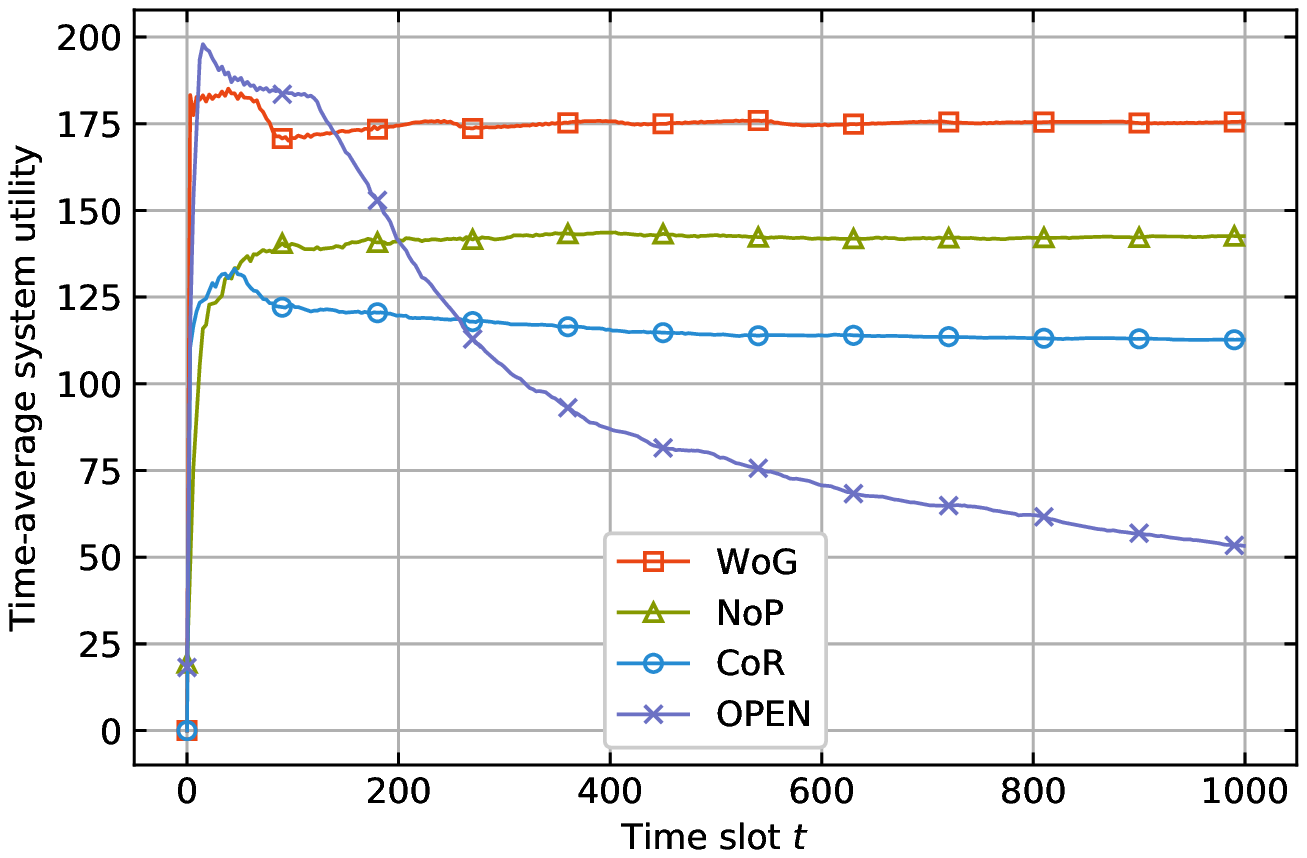} \label{heavy_load_utility}}
\caption{Time-average performance in the heavily loaded case.}
\label{heavy_load}
\end{figure*}

\begin{figure*}[!t]
\begin{minipage}[t]{0.49\linewidth}
\centering
\subfloat[Response Time]{\includegraphics[width=1.7in]{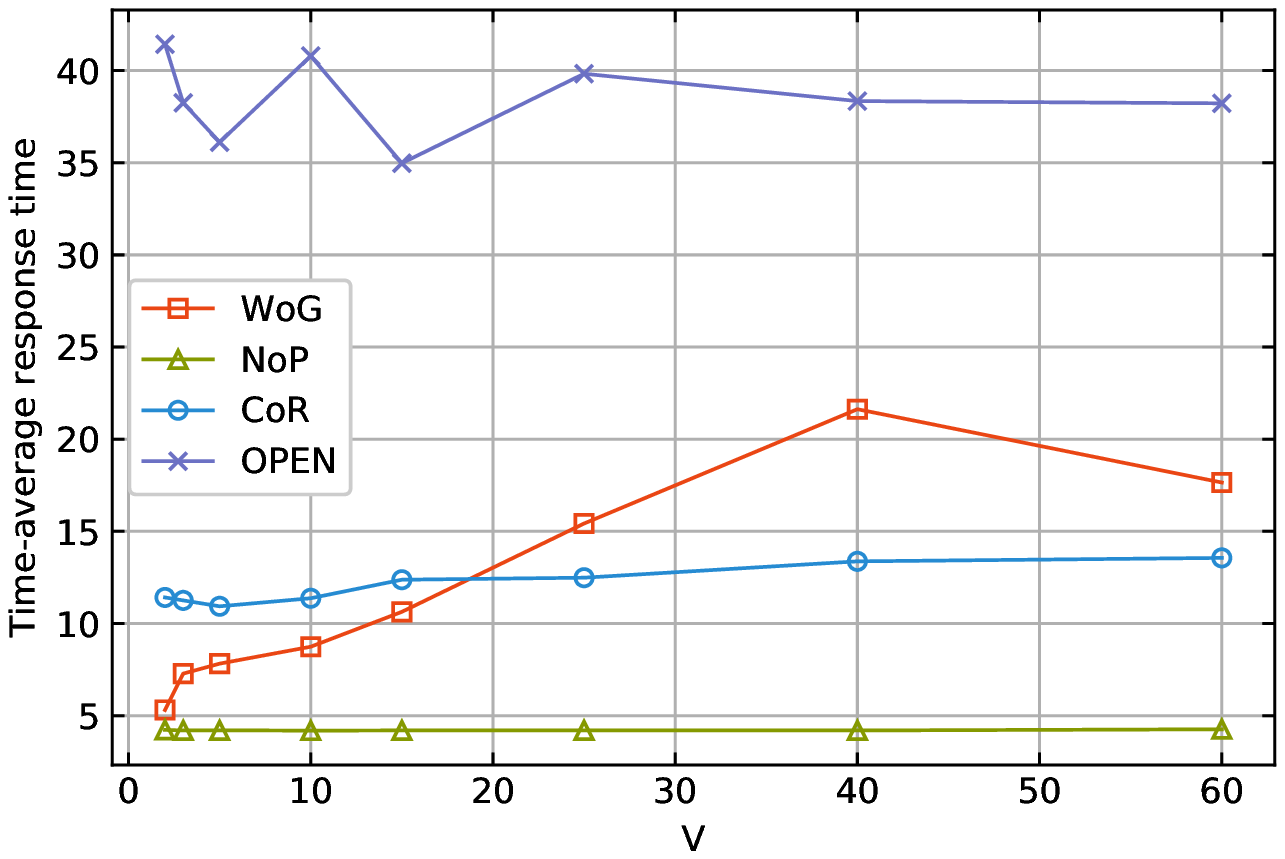} \label{V_delay_heavy}}
\hfil
\subfloat[System Utility]{\includegraphics[width=1.7in]{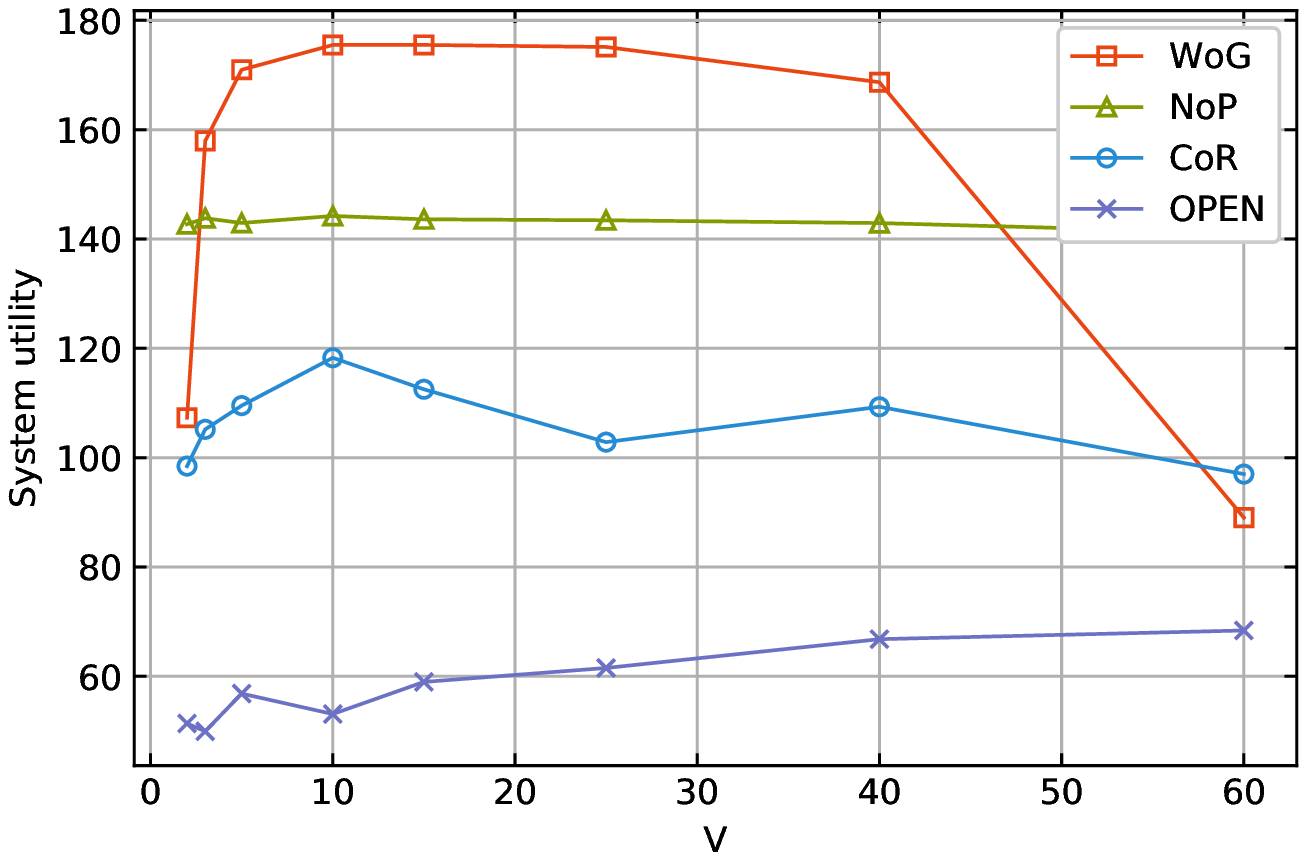} \label{V_utility_heavy}}
\caption{Impact of parameter V in the heavily loaded case.}
\label{diff_V_heavy}
\end{minipage}
\begin{minipage}[t]{0.49\linewidth}
\centering
\subfloat[Response Time]{\includegraphics[width=1.7in]{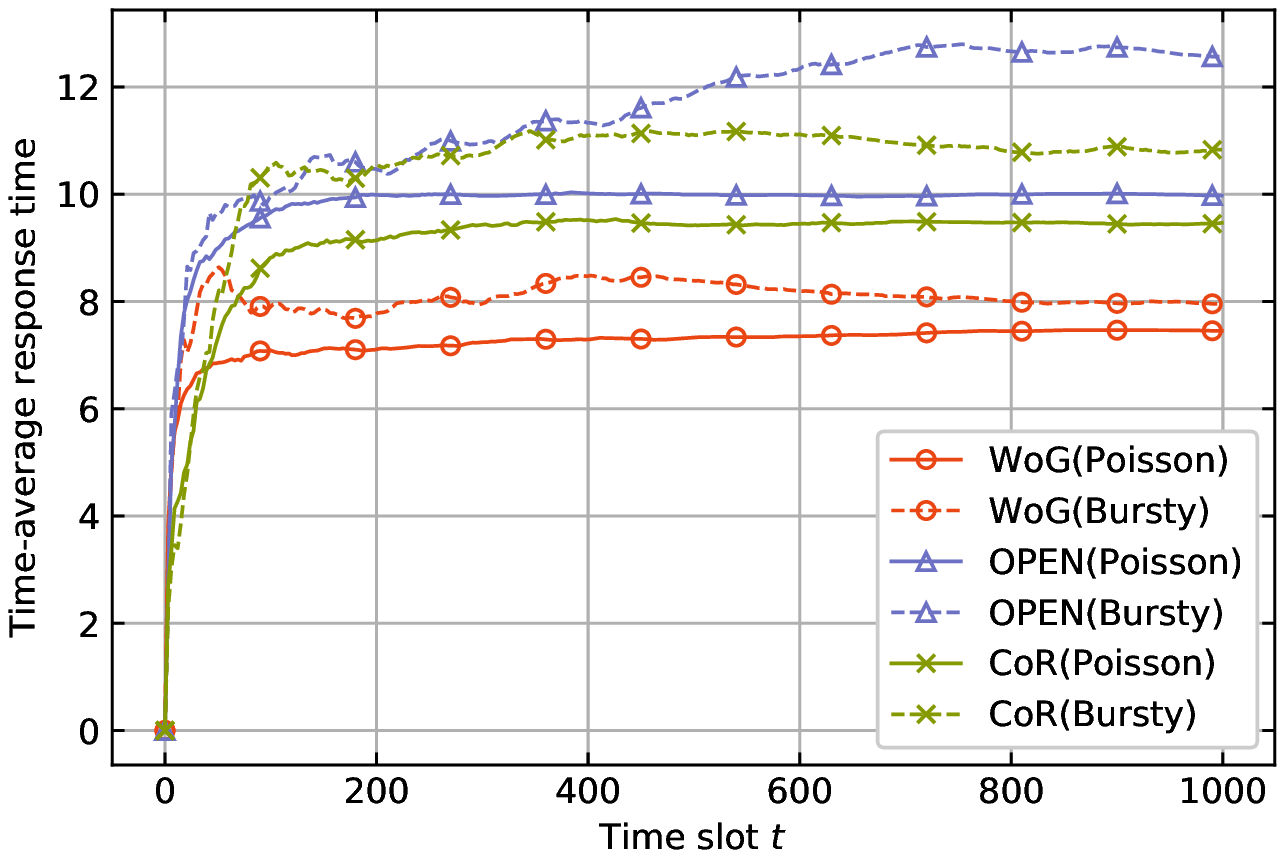} \label{bursty_delay}}
\hfil
\subfloat[System Utility]{\includegraphics[width=1.7in]{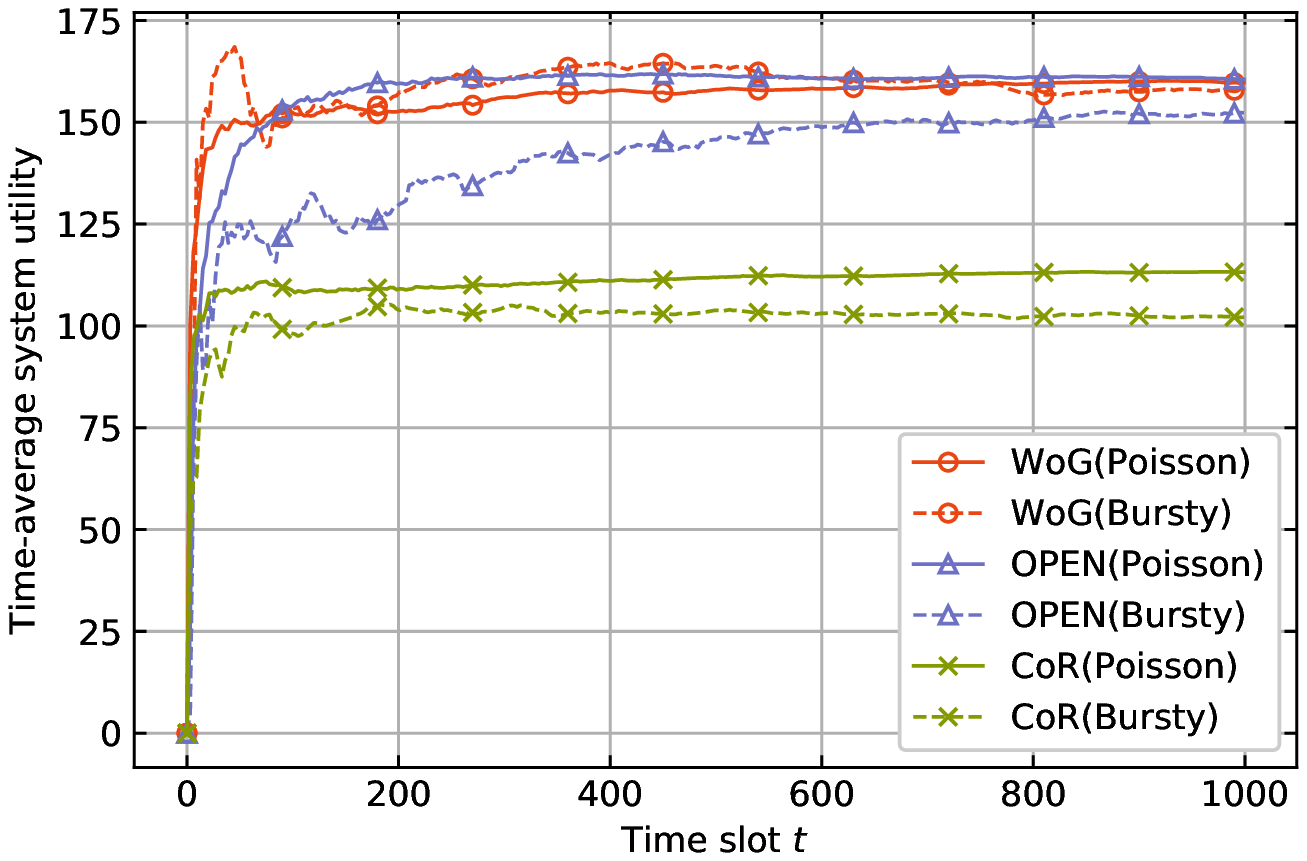} \label{bursty_utility}}
\caption{Impact of task arrival pattern.}
\label{bursty}
\end{minipage}
\end{figure*}

In this section, we evaluate our algorithm presented in Section \ref{section:unknown} under various settings.
The experiments are conducted based on the real-world locations of BSs and end-users within the Central Business District of Melbourne in Australia\footnote{https://github.com/swinedge/eua-dataset}.
We select 36 BSs and 126 user groups whose latitude and longitude lies in $[-37.818166, -37.814257]$ and $[144.958295, 144.966824]$ respectively.
For an arbitrary user group, a task request is generated by a Poisson process with a rate of 0.25 task/ms.
Task requests are submitted to a random BS within 100 meters.
The CPU cycles required by each task are drawn uniformly from [2.5M, 7.5M], and the CPU frequency of each BS is 20GHz \cite{alameddine2019dynamic}.
The energy consumption of one CPU cycle is 8.2nJ, with static energy consumption and long-term energy constraint be 10Wh and 50Wh per hour for each BS.
As in \cite{lyu2018distributed}, the marginal benefit of serving one task is denoted as unit one, so the utility function of each BS is $g_n(x)=x$.
When computing system utility, we also introduce a double punishment for each task that failed to meet the worst-case response time requirement $L^{max}$, which is 50ms in our experiments.
The one-way trip time between BSs is decided by their geographical distances.
Let $Dist_{mn}$ be the distance of BS m and BS n, then $\delta_{mn} = 3ms$ if $Dist_{mn}\in [0m,300m]$, $\delta_{mn}=4ms$ if $Dist_{mn}\in [300m, 600m]$, and $\delta_{mn}=5ms$ if $Dist_{mn} \in [600m, 900m]$.
The tunable parameter $V$ is set to $10$ unless otherwise specified and each slot lasts for 1ms.
The number of algorithm instances $K$ is set to $5$ to deal with the varying workload.

We implement our algorithm that provides Worst-case response time Guarantees (WoG) for 1000 slots and compares it with three benchmarks: (1) No Peer Offloading (NoP): each BS process their own tasks received from end-users and tasks beyond their computing capacity will be blocked; (2) Online Peer Offloading (OPEN) \cite{chen2017computation}: an online peer offloading strategy aiming to minimize the average response time of tasks; (3) Optimization of Collaborative Regions (CoR) \cite{lyu2018distributed}: a cost-effective algorithm that optimizes system utility by maximizing throughput and minimizing average response time.

\subsection{Run-time Performance}
Fig. \ref{light_load} presents the performance comparison of time-average response time and system utility in terms of time slots.
Among the four algorithms, NoP achieves the lowest time-average response time because it blocks tasks for each BS that exceed their computing capacity, and serve the rest tasks as soon as possible.
The side effect is, the system utility of NoP is relatively small due to blocked tasks.
Except for NoP, our algorithm WoG has the lowest time-average response time and obtains the highest system utility together with OPEN.
The performance of CoR seems poor in both metrics.
We found that the scheduling policy of CoR will delay the process of some tasks when the arrived workload of BSs differs significantly.
Besides, the accept decisions of CoR are relatively conservative when the value of $V$ is small and thus cause unnecessary blocks.
We will show later that the utility of CoR is improved with a larger $V$.

\subsection{Impact of V}
We next show the time-average latency and system utility in terms of the tunable parameter V.
The performance of NoP is not affected by V and is regarded as a baseline.
Different from WoG and CoR, the objective of OPEN is response time instead of system utility, so its response time
decreases as V become larger.
As predicted by the theoretical analysis, the response time and system utility of WoG and CoR grow with the increase of V.
The difference is, when V is large, the latency of CoR keeps growing and
cause a reduction of system utility while the performance of WoG is stabilized.
This is because the computing capacity of BSs is adequate in our situation,
so the length of $W_n(t)$ is kept small and encourages BSs to process tasks without further waiting.
Therefore, the average response time remains unchanged when V is large enough.
It should be noted that the results are very different when BSs are overloaded, as shown in the next subsection.

\subsection{Heavily Loaded Case}
In Fig. \ref{heavy_load}, we consider a heavily loaded case where the arrival rate of each user group is enhanced by $50\%$.
In this situation, the average arrived workload will exceed the computing capacity of the whole system.
Fig. \ref{heavy_load_block_rate} and Fig. \ref{heavy_load_satis_ratio} illustrate the time-average block rate and satisfaction ratio of different algorithms, where the satisfaction ratio is defined as the proportion of accepted tasks that are served within $L^{max}=50ms$.
Since OPEN does not block any tasks, its satisfaction ratio drops very quickly as BSs become overloaded and result in a poor system utility.
Although CoR blocks more tasks than WoG, its satisfaction ratio is lower than WoG.
This is because the process of some tasks is delayed in CoR (as mentioned in the previous subsection) and makes them fail to meet the worst-case response time requirement.
The combined effect of block rate and satisfaction ratio is reflected by the time-average system utility given in Fig. \ref{heavy_load_utility}.
We can see that our algorithm WoG achieves the highest utility by blocking tasks as less as possible while maintaining the satisfaction ratio close to $100\%$.
The time-average response time of each algorithm is given in Fig. \ref{heavy_load_delay}.
Not surprisingly, NoP and OPEN have the lowest and highest average latency.
With a higher block rate, the tasks served by CoR are fewer than WoG, thus yield a lower average latency.

The performance under different V in the heavily loaded case is given in Fig. \ref{diff_V_heavy}.
Recall that to ensure the worst-case response time requirement, the value of V should satisfy $\lceil V\max_n\{\nu_n\} \rceil + 2 \leq L^{max} - 2\delta^{max}$, where $\delta^{max} = 5$, $\nu_n = 1$, and $L^{max} = 50$ in our experiments.
Thus, V should be less than $38$.
As a result, the system utility of WoG drops sharply when V exceeds $40$ due to the decrease in satisfaction ratio.
The response time of the rest tasks is improved as there are fewer tasks to be served.
Combining with Fig. \ref{diff_V}, we can see that our algorithm performs well both in light loaded and heavily loaded case
if we have chosen a proper value for V (e.g. $V=10$).

\subsection{Impact of Task Arrival Pattern}
In practice, the real-world task arrival may not follow the assumed Poisson process.
To analyze the practicality of our algorithm, we conduct experiments with different task arrival realizations.
Fig. \ref{bursty} compares the performances of peer offloading algorithms under Poisson and bursty task arrival,
where the latter is implemented with a Markovian arrival process.
We can see that the average response time of all algorithms is degraded but WoG has the smallest increase and still outperforms the others.
In terms of system utility, WoG performs almost equally in both cases.
In contrast, the achieved utility of OPEN and CoR is reduced when dealing with bursty arrivals.
We also run this experiment in the heavily loaded case and observe a similar phenomenon,
which demonstrates the robustness of our algorithm under various task arrival patterns.

\section{Conclusion}
\label{section:conclusion}
In this paper, we studied peer offloading among local BSs with worst-case response time constraint.
We proposed two algorithms for cases with and without the prior knowledge of task arrival rate.
Both the theoretical analysis and numerical results showed our algorithms produce close to optimal performance under strict 
worst-case response time requirement.
One limitation of our work is we can only provide a uniform response time guarantee for all tasks.
More flexible deadlines will be considered in our future work.


%

\appendices
\section{}
\label{appendix:1}
\begin{IEEEproof}
	Using the fact that $\max[a,0]^2 \leq a^2$, we can expand $Z_n(t+1)^2$ and summing over $n\in {1,2,\dots,N}$ 
	\begin{align*}
		\frac{1}{2}\sum_{n=1}^{N}[Z_n(t+1&)^2 - Z_n(t)^2] \leq \frac{1}{2}\sum_{n=1}^{N}(\gamma_n(t) + D_n(t) - \lambda_n)^2 \\
		& - \sum_{n=1}^{N}Z_n(t)[\lambda_n - D_n(t) - \gamma_n(t)].
	\end{align*}
    Apply similar manipulation to $W_n(t)$ and $H_n(t)$.
    Substituting them into \eqref{expr:drift} we have
	\begin{align}
        \Delta&(\bm{\Theta}(t)) \leq \mathbb{E}[B(t)|\bm{\Theta}(t)] - \sum_{n}W_n(t)[E^{aver}_n-e_n(t)|\bm{\Theta}(t)] \notag \\
        & - \sum_{n}Z_n(t)\mathbb{E}[\lambda_n-D_n(t)-\gamma_n(t)|\bm{\Theta}(t)] \notag \\
        & - \sum_{n}\lambda_nH_n(t)\mathbb{E}[(\eta_n(t)+D_n(t))T_n(t)-1|\bm{\Theta}(t)] \label{expr:drift_bound_1}
	\end{align}
	where $B(t)$ is the sum of rest terms.
	Let $\chi(t)$ denote $[\bm{\Theta}(t);\eta_n(t)+D_n(t)]$. 
	Note that by the independence property, if $H_n(t) > 0$, then $T_n(t)$ is independent of 
    $\chi(t)$, so we have $\mathbb{E}[T_n(t)|\chi(t)] = 1/\lambda_n$. 
    Then, by using the law of iterated expectations, we have for 
	any $t$ and $n$ such that $H_n(t)>0$
	\begin{align*}
        \mathbb{E}[(&\eta_n(t)+D_n(t))T_n(t)|\bm{\Theta}(t)] \\
        & = \mathbb{E}[\mathbb{E}[(\eta_n(t)+D_n(t))T_n(t)|\chi(t)]|\bm{\Theta}(t)] \\
        & = \mathbb{E}[(\eta_n(t)+D_n(t))\mathbb{E}[T_n(t)|\chi(t)]|\bm{\Theta}(t)] \\
        & = (1/\lambda_n)\mathbb{E}[(\eta_n(t)+D_n(t))|\bm{\Theta}(t)]
	\end{align*}
    Thus, for any slot $t$ and any BS $n$, we have
	\begin{align}
        \lambda_nH_n(t)\mathbb{E}[&(\eta_n(t)+D_n(t))T_n(t)-1|\bm{\Theta}(t)] \notag \\
        & = H_n(t)\mathbb{E}[\eta_n(t)+D_n(t)-\lambda_n|\bm{\Theta}(t)].\label{expr:lambda_h}
	\end{align}
    The inequality \eqref{lemma1-eq1} follows by substituting \eqref{expr:lambda_h} into the last term of the \eqref{expr:drift_bound_1} and subtracting
    $V\mathbb{E}[\widehat{g}(\bm{\gamma}(t))|\bm{\Theta}(t)]$ from both sides.
    Now we need only to show that $\mathbb{E}[B(t)|\bm{\Theta}(t)] \leq B$ for some finite constant $B$.
    This can be proved by noting that all variables are bounded and $A_n(t)$ is independent of $\bm{\Theta}(t)$.
\end{IEEEproof}

\section{}
\label{appendix:2}
\begin{lemma}
	If $Z_n(t) > V\nu_n$ for some particular $t$ and $n$, then in the first stage of the algorithm we have $\gamma_n(t) = -1$.
	\label{lemma:stage1}
\end{lemma}

This comes easily from the properties of concave functions. Now we can prove the bound of queues
\begin{IEEEproof}
	We first prove by induction that $Z_n(t) \leq \lceil V\nu_n \rceil + 2$ for all $t\geq 0$ and any $n\in \{1,\dots,N\}$.
	If $t=0$, the inequality apparently hold. Suppose the inequality holds at $t$.
	From the update of $Z_n(t)$ we know that $Z_n(t)$ can at most increase by $2$ in every slot.
	So if $Z_n(t) \leq \lceil V\nu_n \rceil$, then $Z_n(t+1) \leq \lceil V\nu_n \rceil + 2$ and the bound holds.
	Else, we have $Z_n(t) > \lceil V\nu_n \rceil$, so $\gamma_n(t) = -1$ by the previous lemma.
	In addition, $D_n(t) \leq 1$ for all slot $t$, so $\gamma_n(t) + D_n(t) \leq 0$, and we have
	$Z_n(t+1) \leq Z_n(t) \leq \lceil V\nu_n \rceil + 2$.
    The bounds of $H_n(t)$ and $W_n(t)$ can be proved similarly.
\end{IEEEproof}
We are left with the proof of the utility bound. We first claim that our constraint $\eta_n(t) + D_n(t) \leq 1$ will not 
affect the optimal value.
\begin{lemma}
	Let $\bm{y}^*$ be the optimal throughput of the relaxed problem with $g^* = g(\bm{y}^*)$.
	Then, there is an algorithm that is independent of $\bm{\Theta}(t)$ 
	and makes randomized decisions that satisfies $\eta_n(t) + D_n(t) \leq 1$ and 
	\begin{equation*}
        \mathbb{E}[\bm{\eta}(t)] = \bm{y}^* \quad \mathbb{E}[\bm{D}(t)] = \bm{\lambda} - \bm{y}^* \quad \mathbb{E}[\bm{e}(t)] = \bm{E}^{max}
	\end{equation*}
    based on the observation of $\bm{A}(t)$.
	\label{lemma:s_only}
\end{lemma}

Please see \cite{neely2013delay} and \cite{neely2009optimal} for a proof. Now we prove \eqref{expr:utility_bound}.
\begin{IEEEproof}
	Since our algorithm satisfies the independence property and minimize the drift-plus-penalty bound, we have
    following inequality by taking expectations of \eqref{lemma1-eq1}
	\begin{align*}
        \mathbb{E}[L&(\bm{\Theta}(t+1))] - \mathbb{E}[L(\bm{\Theta}(t))] - V\mathbb{E}[\widehat{g}(\bm{\gamma}(t))] \\
        \leq & B - V\mathbb{E}[\widehat{g}(\bm{\gamma}^*(t))] - \sum_{n}\mathbb{E}[W_n(t)]\mathbb{E}[E^{aver}_n - e_n^*(t)] \\
        & - \sum_{n}\mathbb{E}[Z_n(t)]\mathbb{E}[\lambda_n - D_n^*(t) - \gamma^*_n(t)] \\
        & - \sum_{n}\mathbb{E}[H_n(t)]\mathbb{E}[\eta^*_n(t) + D^*_n(t) - \lambda_n]
	\end{align*}
    where $\bm{\gamma}^*(t) = \bm{y}^*$, and $\bm{D}^*$, $\bm{\eta}^*$, $\bm{e}^*$ are chosen as in Lemma \ref{lemma:s_only}.
	Plugging into the above formula we have
	\begin{align*}
        \mathbb{E}[L(\bm{\Theta}(t+1))] - \mathbb{E}[L(\bm{\Theta}(t))] - V\mathbb{E}[\widehat{g}(\bm{\gamma}(t))]
		\leq  B - Vg^*.
	\end{align*}
	Summing over $\tau\in \{ 0,\dots,t-1\}$ and dividing by $t$
	\begin{equation*}
        \frac{\mathbb{E}[L(\bm{\Theta}(t))] - \mathbb{E}[L(\bm{\Theta}(0))]}{t} - \frac{V}{t}\sum_{\tau=0}^{t-1}\mathbb{E}[\widehat{g}(\bm{\gamma}(\tau))] 
		\leq B - Vg^*.
	\end{equation*}
	Using the fact that $L(\cdot) \geq 0$ and Jensen's inequality yields
	\begin{equation}
        \widehat{g}(\widebar{\bm{\gamma}}(t)) \geq g^* - B/V - \frac{\mathbb{E}[L(\bm{\Theta}(0))]}{Vt}
		\label{expr:g_hat}
	\end{equation}
    where $\widebar{\bm{\gamma}}(t) \triangleq \frac{1}{t}\sum_{\tau=0}^{t-1}\mathbb{E}[\bm{\gamma}(\tau)]$.
	However, because $Z_n(t)\leq H^{max}_n$, from \eqref{expr:z_virtual_queue} we have
	\begin{equation*}
		\widebar{\bm{y}}(t) + \bm{H}^{max}/t \geq \widebar{\bm{\gamma}}(t)
	\end{equation*}
	where $\bm{H}^{max} = (H^{max}_n)_{n\in \{1,\dots,N\}}$. For all $t$, $-\bm{1} \leq \widebar{\bm{\gamma}}(t) \leq \bm{1}$
	and $\bm{0} \leq \widebar{\bm{y}}(t) \leq \bm{1}$. Therefore
	\begin{equation*}
		[\widebar{\bm{y}}(t) + \bm{H}^{max}/t]^1_0 \geq \widebar{\bm{\gamma}}(t).
	\end{equation*}
	Plugging into \eqref{expr:g_hat} and using the fact that $\widehat{g}$ is non-decreasing
	\begin{equation*}
        \widehat{g}\left( [\widebar{\bm{y}}(t) + \bm{H}^{max}/t]^1_0 \right) \geq g^* - B/V - \frac{\mathbb{E}[L(\bm{\Theta}(0))]}{Vt}.
	\end{equation*}
	By continuity of $\widehat{g}$ and the facts that $\bm{0} \leq \widebar{\bm{y}}(t) \leq \bm{1}$ and $H^{max}_n/t \to 0$
	\begin{equation}
		\liminf_{t\to \infty} \widehat{g}(\widebar{\bm{y}}(t)) \geq g^* - B/V.
		\label{expr:g_hat_bound}
	\end{equation}
    Because $g(\bm{y}) = \widehat{g}(\bm{y})$ when $\bm{0} \leq \bm{y} \leq \bm{1}$, we have \eqref{expr:utility_bound}.
\end{IEEEproof}

\section{}
\label{appendix:3}
\begin{IEEEproof}
    Considering an arbitrary edge region $[t_0, t_1]$ and suppose it contains a violation region $[t'_0, t'_1]$.
    From the proof of Theorem \ref{theorem:performance} we can deduce that the bound for $Z_n(t)$ still holds even if
    the maximum arrival assumption is not satisfied.
    Therefore, $H_n(t) \geq Z_n(t)$ for all $t\in [t_0, t_1]$.
    By the third step of our algorithm, BS $n$ either processes tasks by itself, or drop head-of-line tasks.
    From Section \ref{subsection:block_decision} we know dropped tasks are actually processed by other BSs and their maximum process capability
    is also $w^{max}$.
    As a result, we can conclude that the backlogs on BS $n$ is served with speed $w^{max}/\mbox{slot}$ when $t\in [t_0, t_1]$.

    Let $A^w_n(t:t')$ denote the total workload arrived on the time interval $[t,t']$.
    Since $H_n(t)$ is the waiting time of the head-of-line task,
    we can prove
    \begin{equation}
        w^{max}\times (t - t_0) \leq A^w_n(t_0-H^{max}_n : t-H^{max}_n-T) \quad \forall t\in [t'_0, t'_1].
        \label{eq:ve_arrive}
    \end{equation}
    Let $A^w_n(t)$ be the arrived workload on slot $t$.
    Define a virtual queue $\widehat{H}_n(t)$ with update rule
    \begin{align*}
        \widehat{H}_n(t+1) = \widehat{H}_n(t) + A^w_n(t-H^{max}_n) - w^{max} \quad \forall t\in[t_0, t_1]
    \end{align*}
    and $\widehat{H}_n(t) = 0$ for all other time slots.
    We can prove $\widehat{H}_n(t)$ is always non-negative.
    For any slot $t\in [t'_0, t'_1]$, we have
    \begin{align*}
        \widehat{H}_n&(t-T) = \\
            &A^w_n(t_0-H^{max}_n : t-H^{max}_n-T) - w^{max}\times (t-t_0-T).
    \end{align*}
    Substituting \eqref{eq:ve_arrive} yields
    \begin{equation*}
        \widehat{H}_n(t-T) \geq w^{max}T \quad \forall t\in[t'_0, t'_1].
    \end{equation*}
    Therefore, if $t\in[t_0, t_1]$ belongs to a violation region, then we can find some $t'$ such that $\widehat{H}_n(t')\geq w^{max}T$.
    Thus
    \begin{equation}
        p_{v|e} \leq Pr\{\widehat{H}_n(t)\geq w^{max}T\}.
        \label{eq:pve}
    \end{equation}
    It is well known that when $N_u$ is large while $p_u$ is small, the binomial distribution converges to a Poisson distribution.
    Thus, we can regard $\widehat{H}_n(t)$ as a $M/D/1$ queue and its stationary probabilities satisfies the following inequality \cite{nakagawa2005series}
    \begin{equation*}
        Pr\{\widehat{H}_n(t)\geq w^{max}T\} \leq
        M(r)/r^T\quad \forall T\geq 3
    \end{equation*}
    where
    $r$ is a tunable parameter and $M(r)$ is a constant determined by $r$.
    The detailed definition of $M(r)$ can be found in \cite{nakagawa2005series}.
    Substituting into \eqref{eq:pve} proves our theorem.
\end{IEEEproof}


\ifCLASSOPTIONcaptionsoff
  \newpage
\fi



\bibliographystyle{IEEEtran}
\bibliography{IEEEabrv,ref}
\end{document}